\documentclass[reprint,aps,prx]{revtex4-2}
\usepackage{graphicx} 
\usepackage{amsmath, amsthm, amssymb}
\usepackage{xcolor}
\usepackage{comment}
\usepackage{hyperref}
\usepackage{braket}

\newcommand{\dOne}{\Delta_1}
\newcommand{\dTwo}{\Delta_2}
\newcommand{\w}{\omega}
\newcommand{\wLR}{\omega_{\delta\Delta}}
\newcommand{\e}{\epsilon}

\begin{document}
\title{
Proximity effect in gap-asymmetric superconducting bilayers \\ and regularization of transition rates
}

\author{Giampiero Marchegiani}
\email{giampiero.marchegiani@tii.ae}
\affiliation{Quantum Research Center, Technology Innovation Institute, Abu Dhabi 9639, United Arab Emirates}

\author{Gianluigi Catelani}
\affiliation{Institute for Theoretical Nanoelectronics (PGI-2),Forschungszentrum J\"ulich, 52428 J\"ulich, Germany}
\affiliation{Quantum Research Center, Technology Innovation Institute, Abu Dhabi 9639, United Arab Emirates}

\date{\today}

\begin{abstract}
The standard mean-field treatment of low-temperature superconductors leads to a square-root divergent density of states at the gap value. This feature can lead to unphysical logarithmic divergences in various quantities, such as currents and qubit transition rates. We revisit their possible regularization based on the proximity effect between two superconducting films with different gaps. We derive analytical approximations for the density of states in each superconducting film. We find that the smearing of the density of states grows with the gap asymmetry. As a concrete example, we discuss the regularization of transition rates in qubits with frequency close to resonance with the gap asymmetry between the two films, and the consequent smoothening of the jump discontinuity in the qubit frequency shift. Our results could also aid the development of superconducting detectors and related devices based on superconducting bilayers.
\end{abstract}

\maketitle

\section{Introduction}
In the mean-field treatment of the Bardeen-Cooper-Schrieffer (BCS) theory of superconductivity~\cite{BCSpaper,tinkham}, the quasiparticle density of states (DoS) $\mathcal{N}_j(\epsilon)=|\Re[\epsilon/\sqrt{\epsilon^2-\Delta_j^2}]|$ is characterized by a square-root singularity at the gap ($\Delta_j$) edge, i.e., $\epsilon=\Delta_j$, where $\epsilon$ is the quasiparticle energy with respect to the chemical potential of superconductor $j$. This singularity leads to  theoretical predictions of (unphysical) divergences in various observables. Calculations at the leading order in the transmission amplitude of tunnel-coupled superconductors represent a standard example: transition rates take the form of a convolution between the two quasiparticle densities of states, displaced by the energy difference $\omega$ between initial and final states and weighted by the corresponding occupation factors~\cite{Ingold1992}; when the energy $\omega$ is resonant with the gap asymmetry $|\dOne-\dTwo|$, the square-root divergences in the DoSs align and so the convolution integral leads to a logarithmic divergence~\cite{Shapiro1962}. This divergence is ubiquitous for superconducting junctions: it affects the non-dissipative Josephson component of the tunneling current for a junction biased at the gap-sum (Riedel peak, which also occurs for equal gaps $\dOne=\dTwo$)~\cite{Barone}, the dynamics of current-biased junctions~\cite{AverinPhysRevResearch3}, as well as thermoelectric properties, such as cooling power~\cite{FRANK1997281, Manninen} or the thermoelectric power~\cite{MarchegianiPRL, MarchegianiPRR} of thermally biased superconducting-insulator-superconducting junctions. Experiments showing a finite width of the Riedel peak~\cite{RiedelExperimental} have stimulated theoretical investigation of this divergence since the seventies, even in connection to the sign of the dissipative phase-dependent component (``$\cos\phi$ problem''~\cite{Barone}) of the quasiparticle current. In Ref.~\cite{Zorin1979}, the authors identify possible mechanisms for the broadening of the Riedel peak, 
such as i) inelastic relaxation in the superconducting electrode, which effectively adds a small imaginary part to the superconducting gap in the DoS, ii) inelastic relaxation processes during the tunneling, and iii) possible inhomogeneity and anisotropy of the electrodes.

In the analysis of experimental current-voltage characteristics of superconducting tunnel junctions, one widely used approach to eliminate such divergences consists in adding a small imaginary part $\gamma$ to the energy in the BCS DoS, $\epsilon\to\epsilon+i\gamma$, a parameter originally introduced by Dynes et al. for strong-coupled superconductors~\cite{DynesPRL41}. This approach regularizes the logarithmic divergences, and can account for finite subgap ($\epsilon<\Delta$) DoS typically detected in measurements based on normal-insulator-superconducting junctions. It has been suggested that this modification is the result of environment-assisted tunneling~\cite{PekolaDynes}. The properties of so-called Dynes superconductors have been explored in a series of recent works~\cite{PhysRevB.94.144508, PhysRevB.96.014509, PhysRevB.97.014517, PhysRevB.102.014508}. Alternative mechanisms that can explain the broadening of the peak in the DoS include scattering by magnetic impurities~\cite{PhysRev.136.A1500}. We classify all the regularization schemes discussed so far as extrinsic, since the deviation of the quasiparticle DoS from the BCS model are related to effects beyond the mean-field treatment or couplings to additional degrees of freedom.

However, the divergences can also be regularized intrinsically, that is, within the mean-field description of BCS superconductors. In other words, the singularities only occur in perturbative calculations and are absent in non-perturbative treatments, as already pointed out in Ref.~\cite{Hasselberg_1974_Renormalization}, which has received little attention. The perturbative nature of the singularity was also highlighted in the scattering formulation of Ref.~\cite{CuevasPRB54}, with a particular focus on the subgap singularities associated with multiple Andreev reflection processes in superconductor-superconductor contacts. In the presence of a finite phase bias between two superconductors, the BCS DoS is modified by the presence of the Andreev bound states~\cite{SaulsAndreev}, excitations with energy smaller than the superconducting gap; in particular, the singularity at the gap edge is replaced by a square root threshold~\cite{Kos2013}. Similarly, a supercurrent induces a finite pair-breaking parameter, formally equivalent to the one due to paramagnetic impurities~\cite{FuldePhysRev137, MakiPhysRev140}, that transforms the square-root divergence into a threshold.

In this work, we focus on an intrinsic regularization for gap asymmetric superconductors: we show how the proximity effect changes the DoS in the superconducting electrodes even in the tunneling regime, eliminating the divergences in the case of asymmetric gaps. The BCS singularity regularization associated with the proximity effect has already been discussed in several works, for both normal-superconducting (NS) contacts~\cite{McMillanPR175, FominovPRB63} and superconducting bilayers~\cite{golubov1993josephson,GolubovPRB51, Brammertz}. However, the explicit connection with the Dynes parameter has been only recently highlighted in NS contacts in the weak-coupling regime~\cite{HosseinkhaniPRB97}. In this respect, we generalize the results of Ref.~\cite{HosseinkhaniPRB97} to include weakly-coupled superconductors and argue to what degree a Dynes-like broadening parameter can be used to model the DoS in a proximitized superconductor, providing analytical approximations. We show that the Dynes expression gives a good approximation for the DoS of the high-gap film for sufficiently large gap asymmetry (except close to the gap) and that the effective Dynes parameter is proportional to the gap asymmetry~\footnote{We remark here that in the presence of proximity effect, the order parameters $\Delta_j$ generally differ from the gap in the density of states~\cite{BelzigQuasiclassical}. For concreteness, with a slight language abuse, below we refer to the difference of the order parameters $\delta\Delta=|\Delta_{1}-\Delta_2|$ as gap asymmetry.}. 
Finally, we discuss the consequence of this regularization in the context of the relaxation rate of a transmon qubit with asymmetric junctions~\cite{MarchegianiPRXQuantum}. 

This work is organized as follows: in Sec.~\ref{sec:Model} we present the theoretical model for the proximity effect and the parametrizations used for analytical calculations. Next, we derive analytical approximation for the two DoS in Sec.~\ref{sec:approxSolutions}. Our results are applied to the concrete case of a transmon qubit in Sec.~\ref{sec:rates}, where we discuss the regularization of the divergence in the relaxation rate due to quasiparticle tunneling and the smoothing of the jump discontinuity in the related frequency shift. We summarize our findings in Sec.~\ref{sec:Conclusions}. 

\begin{figure}
    \centering
\includegraphics{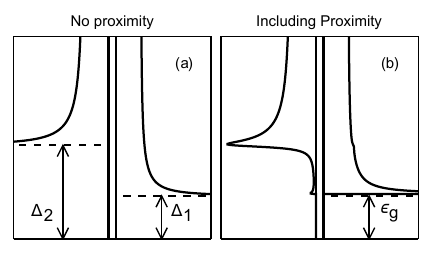}
    \caption{Band schematic for superconducting films. (a)
    Zero transparency (no proximity effect): the BCS quasiparticle DoSs are characterized by square-root singularities at the gap edge. (b) Small transparency (weak proximity): the gap $\epsilon_g\gtrsim\dOne$ globally characterizes the full structure, with finite DoS in the left electrode for $\epsilon_g<\epsilon<\dTwo$. At the gap edge, the square root singularity is replaced by a square-root threshold (see discussion in the text), and each peak is broadened.
    }
    \label{fig:Figure1}
\end{figure}
\section{Model}
\label{sec:Model}

We consider a quasi-classical description~\cite{RammerRMP58,BelzigQuasiclassical} of a bilayer composed of two superconducting films (denoted as $\mathrm{S}_1$ and $\mathrm{S}_2$), neglecting modulation of the spectral properties on the scale of the Fermi wavelength. In thermal equilibrium, such properties are encoded in the retarded Green's function, a $2\times2$ matrix in the electron-hole space (we consider standard s-wave superconductors and neglect spin-dependent tunneling). For dirty films, characterized by a mean free path shorter than the superconducting coherence length, the Green's function satisfies the Usadel equation~\cite{Usadel}. Unless explicitly stated, we work in units $\hbar=k_B=1$, where $k_B$ is the Boltzmann constant and $\hbar$ is the reduced Planck's constant. To illustrate the effect of tunneling on the spectral properties of the superconducting films, we consider a uniform bilayer with film thicknesses small compared to respective coherence lengths, so that the Green's functions in each layer are independent of position. In Sec.~\ref{sec:rates} we discuss the relevance of our modeling to superconducting qubits. The derivation of the Usadel equations for normal-superconducting bilayers has been presented in detail in Refs.~\cite{FominovPRB63} and \cite{HosseinkhaniPRB97}. Extending those derivations to the case of two superconducting films and
assuming no phase difference between the superconductors, the equations read:
\begin{equation}
i\epsilon \sin\theta_j(\epsilon)+\Delta_j(T)\cos\theta_j(\epsilon)=\frac{1}{\tau_j}\sin[\theta_j(\epsilon)-\theta_{\tilde{j}}(\epsilon)]\, ,
\label{eq:UsadelAngular}
\end{equation}
where $\tilde j = 2$ for $j=1$ and \textit{vice versa}. 
The times $\tau_j = 2e^2 \nu_{0,j} d_j R_{\rm int} \mathcal{A}$  are proportional to the $\mathrm{S}_1\mathrm{S}_2$ interface resistance times area product $R_{\rm int} \mathcal{A}$ and the density
of states at the Fermi level $\nu_{0,j}$ of the two films. In deriving Eq.~\eqref{eq:UsadelAngular}, we assumed the Kupriyanov-Lukichev boundary conditions at the $\mathrm{S}_1\mathrm{S}_2$ interface~\cite{Kupriyanov}, valid for small transparencies. The angle-variables $\theta_j(\epsilon)$ are generally complex-valued; the density of states in the film $j$ is given by 
\begin{equation}
\mathcal{N}_j(\epsilon)=\Re[\cos\theta_j(\epsilon)] \, ,
\label{eq:DoSGreen}
\end{equation}
where $\Re[z]$ denotes the real part of $z$. We note that Eq.~\eqref{eq:UsadelAngular} coincides with the McMillan tunneling model~\cite{McMillanPR175} of the superconducting proximity effect~\footnote{In the notation of Ref.~\cite{McMillanPR175}, $\tau_j=1/\Gamma_j$, $\cos\theta_j(\e)=-i\epsilon/\sqrt{\Delta_j^2(\e)-\epsilon^2}$, $\sin\theta_j=\Delta_j(\e)/\sqrt{\Delta_j^2(\e)-\epsilon^2}$ and $\Delta_j=\Delta_j^{\rm ph}$, where $j=1,2$ correspond to $j=N,S$ in McMillan's work. The equivalence between the McMillan model and the quasiclassical description has also been extensively discussed in Refs.~\cite{GolubovPRB51,golubov1993josephson}}. The order parameter in each layer must be calculated self-consistently, requiring it to satisfy the following equation
\begin{equation}
\Delta_j(T)=\frac{\nu_{0,j}\lambda_j}{2}\int_{-\w_{D,j}}^{\w_{D,j}}d\e\tanh\left(\frac{\epsilon}{2T}\right)
p_j(\e)\, ,
\label{eq:selfGap}
\end{equation}
where $p_j(\e)=\Im[\sin\theta_j]$ is the pair correlation function in the film $j$, $\Im[z]$ denotes the imaginary part of $z$, and both superconductors are assumed in thermal equilibrium with quasiparticle temperature $T$. In Eq.~\eqref{eq:selfGap}, $\lambda_j$ and $\omega_{D,j}$ are the pairing coupling strength and the Debye energy of superconductor $j$, respectively. For notational simplicity, hereinafter we omit the temperature dependence of the order parameters. For our goals, we only consider temperatures much smaller than the gap $\epsilon_g$ in the quasiparticle spectrum; then this dependence is exponentially small in $T/\epsilon_g$ and can be disregarded.

This work's main purpose is to estimate the modification to the superconducting DoS due to the proximity effect in the tunneling limit in which the two films are weakly coupled, $\tau_j\Delta_j\gg 1$. When this coupling is neglected, the right-hand side in Eq.~\eqref{eq:UsadelAngular} is zero; the pairing angles read $\theta_j(\e)=\arctan[i\Delta_j/\e]$. The quasiparticle DoS coincides with the standard result of the BCS model, characterized by a square root singularity at $\e=\Delta_j$ [cf.~Fig.~\ref{fig:Figure1}(a)]. The situation is unchanged for $\dOne=\dTwo$ and arbitrary couplings $\tau_j$: indeed, by symmetry, $\theta_1(\e)=\theta_2(\e)$ and the right-hand side of Eq.~\eqref{eq:UsadelAngular} vanishes, thus reducing to the uncoupled situation. Next, we consider, without loss of generality, $\dOne<\dTwo$: for $\dOne= 0$, i.e., normal-superconducting bilayer, it has been theoretically shown that the coupling leads to a finite subgap DoS in the superconductor, for $\e_g<\e<\dTwo$, similar to the effect of a finite lifetime approximately equal to $\tau_2$~\cite{HosseinkhaniPRB97}. In particular, for finite coupling, the square-root singularity is replaced by a square-root threshold around the minigap $\e_g\simeq \tau_1^{-1}$. In a similar fashion, we show below for a superconducting bilayer that the BCS singularity becomes a square-root threshold at the gap $\e_g\gtrsim \Delta_1$, and correspondingly, a small DoS appears in the higher-gap superconductor for energies larger than $\e_g$ [cf.~Fig.~\ref{fig:Figure1}(b)]. The smearing of the BCS singularities due to the proximity effect can be qualitatively understood as follows. When the two films are coupled, electrons can tunnel between them (on timescales characterized by the quantities $\tau_1$ and $\tau_2$). As a result, the quasiparticle states in each film are no longer eigenstates of the bilayer and so they acquire a finite lifetime; this lifetime, in turn, produces the smearing of the singularities.

\subsection{Parametrization of the Usadel equations}
\label{sec:Usadel_param}
To derive approximate solutions to the Usadel equations (see Sec.~\ref {sec:approxSolutions}), it is convenient to change the parametrization, setting $\theta_j=\pi/2+i\chi_j$. Defining $X_j=\sinh[\chi_j]$, the Usadel equations read
\begin{equation}
\label{eq:usadelXj}
\e \sqrt{1+X_j^2}-\Delta_j X_j =\frac{1}{\tau_j}\left[X_j\sqrt{1+X_{\tilde{j}}^2}-X_{\tilde{j}}\sqrt{1+X_j^2}\right].
\end{equation}
We recall that $X_j$ is generally a complex variable so the square root may comprise both possible roots. The physically significant solution is the one with positive DoS; the DoS is given by the imaginary part of $X_j$, i.e., $\mathcal{N}_j(\epsilon)=\Im[X_j(\epsilon)]$, and similarly the pair-correlation function reads $p_j(\epsilon)=\Im[\sqrt{1+X_j^2(\epsilon)}]$ (which must be odd in $\e$). 
To facilitate the comparison with previous works on superconducting bilayers~\cite{golubov1989,golubov1993josephson,GolubovPRB51,Brammertz}, we note that $X_j$ and $\sqrt{1+X_j^2}$ correspond to the normal and anomalous Green's function in superconductor $j$, up to a multiplicative factor given by the imaginary units $i$~\footnote{Recall also that the Matsubara Green's function can be obtained with the substitution $\epsilon\to i\omega_k$, with $\w_k=2\pi T(k +1/2)$ and $k\in \mathbb{N}$.}.

Using Eq.~\eqref{eq:usadelXj}, we can express $X_{j}$ and $\sqrt{1+X_j^2}$ as a function of $X_{\tilde{j}}$
\begin{align}
X_j &=\! \frac{\tau_j\epsilon + X_{\tilde{j}}}{\sqrt{1+\tau_j^2(\Delta_j^2-\e^2)+2\tau_j[\Delta_j\sqrt{X_{\tilde{j}}^2+1}-\e X_{\tilde{j}}]}},
\label{eq:XjVsXjtilde}\\
\sqrt{1+X_j^2} &=\! \frac{\tau_j\Delta_j + \sqrt{1+X_{\tilde{j}}^2}}{\sqrt{1+\tau_j^2(\Delta_j^2-\e^2)+2\tau_j[\Delta_j\sqrt{X_{\tilde{j}}^2+1}-\e X_{\tilde{j}}]}},
\label{eq:XjsqVsXjtilde}
\end{align}
and write a single-variable equation 
\begin{align}
\frac{\tau_{\tilde{j}}\Delta_{\tilde{j}}}{\tau_j\e}&\left(X_{\tilde{j}}-\frac{\e}{\Delta_{\tilde{j}}}\sqrt{X_{\tilde{j}}^2+1}\right)
\nonumber\\
&\times\sqrt{1+\tau_j^2(\Delta_j^2-\e^2)+2\tau_j[\Delta_j\sqrt{X_{\tilde{j}}^2+1}-\e X_{\tilde{j}}]}
\nonumber\\
&=\sqrt{X_{\tilde{j}}^2+1}-\frac{\Delta_j}{\e}X_{\tilde{j}}\,.
\label{eq:selfXj}
\end{align}
For $\Delta_j=0$ (NS bilayer), Eq.~\eqref{eq:selfXj} reduces to Eq.~(B3) of Ref.~\cite{HosseinkhaniPRB97} identifying $\dTwo=\Delta$, $\tau_1=\tau_N$, and $\tau_2=\tau_S$.  

\subsection{Dynes parametrization of the Usadel equation}
\label{sec:Dynes_param}

An alternative parametrization can be obtained from the one presented in Sec.~\ref{sec:Usadel_param} by writing $X_j$ in the form 
\begin{equation}
\label{eq:DynesAnsatz}
X_j(\e)= \frac{\e+i\gamma_j(\e)}{\sqrt{\Delta_j^2-[\e+i\gamma_j(\e)]^2}} \, ,
\end{equation}
By plugging Eq.~\eqref{eq:DynesAnsatz} into Eq.~\eqref{eq:selfXj}, we can express the Usadel equation as
\begin{equation}
\label{eq:iterativeFormGammaj}
i\gamma_j=\frac{\tau_{\tilde j}\e(\Delta_j-\Delta_{\tilde j})}{\tau_{\tilde j}\Delta_{\tilde j}+\tau_j\Delta_j\sqrt{\mathcal{G}_j(\e;i\gamma_j)}}\,,
\end{equation}
where, for notational convenience, we introduce the function 
\begin{equation}
\label{eq:squareRootGammaj}
\mathcal{G}_j(\e;z)=\tau_{\tilde j}^2(\Delta_{\tilde j}^2-\e^2)+1+2\tau_{\tilde j
}\frac{\dOne\dTwo-\e(\e+z)}{\sqrt{\Delta_j^2-(\e +z)^2}}\, .
\end{equation}
Equation~\eqref{eq:iterativeFormGammaj} can be
used to compute the numerical root iteratively by choosing an initial condition for the complex variable $z = i\gamma_j$ (see Appendix~\ref{app:Dynes}).

\section{Approximate solutions for uniform bilayer}
\label{sec:approxSolutions}
Equation~\eqref{eq:selfXj} is a polynomial expression in $X_j$ of order six~\footnote{This result is in agreement with the one reported for NS bilayers in Ref.~\cite{FominovPRB63}, where the authors use a different variable, $Z=\exp(i\theta_S)$.}, and thus its solution cannot be expressed as an explicit combination of radicals of the polynomial coefficients. Nevertheless, the solution can be obtained numerically with a standard find-root procedure, selecting the proper root to ensure that the density of states is positive above the gap (and so the pair correlation function for $\e>0$). The purpose of this section is to obtain explicit analytical approximations for $X_j$ and consequently for the DoS in each film. These expressions are exploited in Sec.~\ref{sec:rates} to  estimate the impact of proximity effect on the quasiparticle transition rates in superconducting qubits. With a slight abuse of language, we use for convenience the terms low-gap and high-gap superconductor to denote $\mathrm{S}_1$ and $\mathrm{S}_2$ respectively, even though the gap characterize the full structure.

\subsection{Subgap DoS and broadening in the high-gap superconductor}
\label{sec:smearingDoSX2}

\begin{figure}
\includegraphics{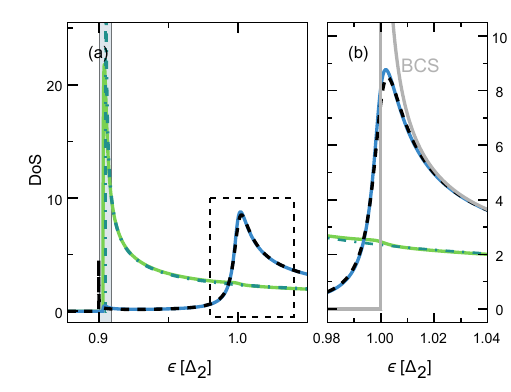}
    \caption{Broadening of the BCS singularity and subgap states in the high-gap superconductor. (a) DoS of $\mathrm{S}_1$ and $\mathrm{S}_2$ computed solving numerically Eq.~\eqref{eq:UsadelAngular} (solid curves). The dashed line is the analytical approximations obtained using Eq.~\eqref{eq:explicitGamma} in Eq.~\eqref{eq:DynesAnsatz} for $\mathrm{S}_2$ and the dot-dashed one a BCS DoS with shifted gap $\tilde{\e}_g$ [Eq.~\eqref{eq:tildeEpsWeak}] for $\mathrm{S}_1$. These approximations are accurate for weakly-coupled superconductors except close to the gap (shaded aquamarine region). (b) Enlarged view for energies close to $\dTwo$ (corresponding to the dashed rectangle in panel a) showing the broadening of the $\mathrm{S}_2$ DoS. The BCS DoS is also plotted for comparison. Parameters: $\tau_1=\tau_2=50/\Delta_2$ and $\dOne=0.9\dTwo$.
    }
    \label{fig:Figure2}
\end{figure}
For uniform NS bilayers, it was shown in Ref.~\cite{HosseinkhaniPRB97} that the superconducting DoS is well approximated by the Dynes formula in the weak coupling regime $\tau_2\dTwo\gg 1$ for energies larger than (and not too close to) the minigap $\epsilon_g\simeq 1/\tau_1$. Analogously, here we show how a finite transparency of the $\mathrm{S}_1\mathrm{S}_2$ barrier determines a non-zero DoS in the high-gap film at subgap energies $\e_g<\e<\dTwo$, where  $\e_g\gtrsim\dOne$ is the DoS gap in the weak coupling limit $\tau_j\Delta_j\gg 1$. To show this feature, we adopt a perturbative approach. Specifically, when the condition  
\begin{equation}
\tau_1^2(\e^2-\dOne^2)\gg \left|1+2\tau_1\left[\dOne\sqrt{X_{2}^2+1}-\e X_{2}\right]\right|\, ,
\label{eq:conditionX2weak}
\end{equation}
is satisfied, we keep only the first term in the right-hand side of Eq.~\eqref{eq:squareRootGammaj} (with $j=2$).
For energies $\e>\dOne$, and selecting in Eq.~\eqref{eq:iterativeFormGammaj} the square root branch with negative imaginary part to ensure a positive-defined DoS, we immediately find
\begin{align}
\gamma_2(\e) &\simeq \frac{\e(\dTwo-\dOne)}{
i\dOne+
\tau_2\dTwo\sqrt{\e^2-\dOne^2}}
\label{eq:explicitGamma}\\
&\simeq 
\frac{\e(\dTwo-\dOne)}{
\tau_2\dTwo\sqrt{\epsilon^2-\dOne^2}}
\, ,
\label{eq:explicitGamma_2nd}
\end{align}
where in the second line we assumed $\tau_2\dTwo\sqrt{\e^2-\dOne^2}\gg \dOne$.
For energies not too close to $\dOne$ -- below we comment quantitatively on this point -- $\gamma_2(\epsilon)$ is a real-valued function and $|\gamma_2(\epsilon)|\ll |\epsilon|$; the latter inequality implies that at these energies the DoS and the pair correlation function are approximately related as in BCS theory, $p_2\simeq \Delta_2 \mathcal{N}_2/\epsilon$. Moreover, for $\epsilon\gg \dOne$,  $\gamma_2(\epsilon)$ approaches the constant value $(\dTwo-\dOne)/(\tau_2\dTwo)$, so that the DoS of the high-gap superconductor is of the Dynes form, generalizing the result obtained in Ref.~\cite{HosseinkhaniPRB97} for NS bilayers (to which our formulas reduce for $\Delta_1=0$)~\footnote{For $\dOne=0$, $\gamma_2=1/\tau_2$ with no extra condition on the energy other than Eq.~\eqref{eq:conditionX2weak}}. In other words, the DoS of the high-gap superconductor in weakly-coupled bilayers is well approximated by a Dynes-like expression for energies sufficiently larger than $\dOne$, but we stress that $\gamma_2$ in general depends on energy. Since $\Re[\gamma_2]\neq 0$, the BCS singularity in the DoS for $\e= \dTwo$ is smeared out; notably, this smearing is proportional to the gap asymmetry $\dTwo-\dOne$, and so it vanishes for $\dOne=\dTwo$ (see also Sec.~\ref{sec:Model}).

Using Eq.~\eqref{eq:explicitGamma}, we can estimate the position and the (finite) value of the maximum DoS in the high-gap superconductor at the leading order in $\tau_2\dTwo\gg 1$. Close to $\epsilon=\dTwo$, we can approximate  $\gamma_2(\e\simeq \dTwo)\approx \gamma_{2,0}=\frac{1}{\tau_2}\sqrt{\frac{\dTwo-\dOne}{\dTwo+\dOne}}$, assuming $\sqrt{\dTwo^2-\dOne^2}\gg \dOne/\tau_2\dTwo$. The energy ($\e_{\rm max}^{\mathrm{S}_2}$) at which the DoS is maximum is shifted compared to the uncoupled case~\cite{DynesNote},
\begin{equation}
\epsilon_{\rm max}^{\mathrm{S}_2}-\dTwo\approx \frac{\gamma_{2,0}}{\sqrt{3}}=\frac{1}{\tau_2\sqrt{3}}\sqrt{\frac{\dTwo-\dOne}{\dOne+\dTwo}}\, ,
\label{eq:positionMaximumN2Weak}
\end{equation}
and the peak value of the DoS in the high-gap film reads approximately (keeping terms at the leading order in $\gamma_{2,0}/\dTwo\ll 1)$
\begin{equation}
\mathcal{N}_2(\epsilon_{\rm max}^{\mathrm{S}_2})\approx \frac{3^{3/4}}{4\sqrt{\gamma_{2,0}/\dTwo}}\, .
\label{eq:MaximumN2Weak}
\end{equation}
In Fig.~\ref{fig:Figure2}(a), we display the DoS of the two films obtained by solving numerically Eq.~\eqref{eq:UsadelAngular} (solid curves). Using the approximation in Eq.~\eqref{eq:explicitGamma}, we obtain an accurate description of the DoS (black dashed curve), except in the proximity of the gap (aquamarine area); we can check the range of validity of our expressions by inserting Eqs.~\eqref{eq:DynesAnsatz} and \eqref{eq:explicitGamma} into Eq.~\eqref{eq:conditionX2weak} [see Appendix~\ref{app:energyRangeWeakDynes} for details], which results in the inequality
\begin{equation}
\label{eq:conditionGammaExplicit}
    \epsilon-\dOne\gg \frac{1}{\tau_1}\sqrt{\frac{\dTwo-\dOne}{\dOne+\dTwo}}
    \, .
\end{equation}
This condition indicates that Eq.~\eqref{eq:explicitGamma} is inaccurate for energies close to the gap (the right-hand side is the lowest order correction to the gap, see the next subsection)~\footnote{As discussed in Appendix~\ref{app:expansionWeak}, it is also assumed that $\sqrt{1-\delta^2}/\delta \gg 1/(\tau_1\dOne)$, so the two gaps cannot be arbitrarily close.}. In fact, Eqs.~\eqref{eq:DynesAnsatz} and \eqref{eq:explicitGamma} incorrectly predict a divergence of the DoS for $\epsilon\to \dOne$. Moreover, the approximate expression in Eq.~\eqref{eq:explicitGamma} is slightly less accurate for energies close to the maximum of the DoS [cf. Eq.~\eqref{eq:positionMaximumN2Weak}], see Fig.~\ref{fig:Figure2}(b) (black dashed curve), where the last term in the right-hand-side of Eq.~\eqref{eq:conditionX2weak} approaches its maximum value. In this region, it is possible to obtain a more accurate approximation with an iterative procedure that starts from the approximate formula in Eq.~\eqref{eq:explicitGamma} (see Appendix~\ref{app:iterativeGamma} for details).

\subsection{DoS in the low-gap film close to the gap}
\label{sec:lowerGapDoS}
We now turn to the DoS in the low-gap film. In the previous subsection, we have derived the approximate Eq.~\eqref{eq:explicitGamma} which is accurate except in a region close to the gap [see Eq.~\eqref{eq:conditionGammaExplicit} and Fig.~\ref{fig:Figure2}(a)]. When Eqs.~\eqref{eq:DynesAnsatz} and \eqref{eq:explicitGamma} are inserted into Eq.~\eqref{eq:XjVsXjtilde}, and under the condition of Eq.~\eqref{eq:conditionX2weak}, one can verify that the solution for $X_1$ at the leading order in $\tau_1\dOne\gg 1$ corresponds to the uncoupled (BCS) result, $X_1\approx \epsilon/{\sqrt{\dOne^2-\epsilon^2}}$; sub-leading order deviations mainly occur for $\epsilon\sim\dTwo$, where $\Im[X_2]\propto \sqrt{\tau_2\dTwo} $ [cf. Eq.~\eqref{eq:MaximumN2Weak}]. The precise characterization of this feature, which appears as a ``shoulder'' in the DoS [slightly visible in Figs.~\ref{fig:Figure1}(b) and \ref{fig:Figure2}(b)], goes beyond the scope of this manuscript, and does not significantly affect the weak-coupling regime. For higher transparencies of the $\mathrm{S}_1\mathrm{S}_2$ interface, a more pronounced structure in $X_1$ near $\Delta_2$ may appear for a thick $\mathrm{S}_2$ film~\cite{GolubovPRB51} [see also Ref.~\cite{Fominov2023}, where the shape of this singularity is analyzed in detail for a NS bilayer]. For energies close to the gap $|\epsilon-\dOne|\lesssim 1/\tau_1$, using Eq.~\eqref{eq:explicitGamma} in Eq.~\eqref{eq:DynesAnsatz} is not justified.
One can repeat for $X_1$ the analysis done for $X_2$ in Sec.~\ref{sec:smearingDoSX2}, using the assumption in Eq.~\eqref{eq:conditionX2weak} (after exchanging the subscripts $1\leftrightarrow2$) to write equations analog to Eqs.~\eqref{eq:DynesAnsatz} and \eqref{eq:explicitGamma} for $X_1$ and $\gamma_1$. For $\epsilon\simeq\dOne$, we can approximate $i\gamma_1(\epsilon)\simeq i\gamma_{1,0}=-\sqrt{(\dTwo-\dOne)/(\dOne+\dTwo)}/\tau_1$; in other words, within this approximation the square-root singularity persists  but is shifted to
\begin{equation}
    \tilde\epsilon_g=\dOne + \frac{1}{\tau_1}\sqrt{\frac{\dTwo-\dOne}{\dOne+\dTwo}}\, .
    \label{eq:tildeEpsWeak} 
\end{equation}
The same correction to the gap position was reported in Ref.~\cite{GolubovPRB51} for a superconducting bilayer composed of a thick layer in contact with a thin film with lower order parameter. Note that because of the shifted BCS form of $X_1$, we have $p_1 \simeq \dOne \mathcal{N}_1/(\e +i\gamma_1)\simeq \dOne \mathcal{N}_1/\e$ and hence $p_1 \simeq \mathcal{N}_1$ near the gap.

The accurate description of the DoS close to the gap requires a more careful analysis, perturbative in the small parameters $1/\tau_j\Delta_j$ and $\epsilon-\tilde\epsilon_g$. For this expansion, where we also assume $(\tau_2\dTwo)^2\gg \tau_1\Delta_1$, it is convenient to consider the variable $u=\sqrt{1+X_1^2}-X_1$, which is small in the vicinity of the gap $u \propto 1/(\tau_1\dOne\tau_2\dTwo)^{1/3}$ (see Appendix~\ref{app:expansionWeak} for details). 
In terms of $u$, the DoS is expressed as $\mathcal{N}_1(\epsilon)=-\Im [u(\epsilon)] [1+|u(\epsilon)|^{-2}]/2\simeq -\Im [u(\epsilon)]/(2|u(\epsilon)|^{2})$; note that the smallness of $u$ implies $p_1 \simeq \mathcal{N}_1$, confirming the validity of this approximate relation near the gap. In the first subleading order, $u$ satisfies a depressed cubic equation
\begin{equation}
u^3+\frac{(\epsilon-\tilde\epsilon_g)}{2\dOne}u + 
\frac{\sqrt{\Delta_2-\Delta_1
}}{4\tau_1\tau_2\dOne(\Delta_2+\Delta_1)^{3/2}}=0 \, ,
\label{eq:depressedCubicWeak}
\end{equation}
where above the gap we select the complex root with negative imaginary part to have $\mathcal{N}_1(\e)\geq 0$. 
At the first subleading order of the perturbative expansion, the DoS is maximum at $\epsilon=\tilde\epsilon_g$, where it approximately reads
\begin{equation}
\mathcal{N}_1(\tilde{\epsilon}_g)\approx \frac{3}{2^{4/3}\sqrt{2}}\sqrt{\frac{\dOne}{\tilde\epsilon_g-\epsilon_g}}\, .
\label{eq:MaximumDosX1Weak}
\end{equation}
where the gap position $\epsilon_g$ (at the same order) satisfies
\begin{equation}\label{eq:minigapWeak}
    \frac{\tilde\epsilon_g-\epsilon_g}{\dOne}=\frac{3}{2(\tau_1\dOne)^{2/3}(\tau_2\dTwo)^{2/3}}\frac{(1-\delta)^{1/3}}{1+\delta}
\end{equation}
with $\delta=\dOne/\dTwo$ (note that the gap position is given in terms of the self-consistent order parameter $\Delta_1$, which is larger than the value $\Delta_{1,0}$ it has in the limit of no proximity, see Appendix~\ref{app:selfconsistence}). Expanding the solution to Eq.~\eqref{eq:depressedCubicWeak} around the gap, it can be shown that the BCS singularity is replaced by a square-root threshold, in agreement with the result of Ref.~\cite{HosseinkhaniPRB97} for NS bilayers. However, while at leading order the value of the minigap in NS bilayers~\cite{McMillanPR175,FominovPRB63,HosseinkhaniPRB97} is recovered by setting $\dOne=0$ in Eq.~\eqref{eq:tildeEpsWeak}, Eq.~\eqref{eq:minigapWeak} does not capture the next-order correction for such systems~\cite{HosseinkhaniPRB97}, since our result is derived under the assumption $\tau_1\dOne\gg 1$. Inserting Eq.~\eqref{eq:minigapWeak} into Eq.~\eqref{eq:MaximumDosX1Weak}, one finds that the maximum DoS approximately reads $\mathcal{N}_1(\tilde{\e}_{g})\simeq \sqrt{3(1+\delta)}(\tau_1\dOne\tau_2\dTwo)^{1/3}/2^{4/3}(1-\delta)^{1/6}$, which diverges for gap-symmetric junctions ($\delta=1$).

Equation~\eqref{eq:minigapWeak} gives an estimate for the broadening of the DoS for $\tau_j\Delta_j\gg 1$,
\begin{equation}
\tilde{\gamma}_1 =\tilde\e_g-\e_g \, .
\label{eq:gamma1Broadening}
\end{equation}
By comparing Eqs.~\eqref{eq:MaximumN2Weak} and \eqref{eq:MaximumDosX1Weak}, at the leading order the maximum value of the DoS would be the same as for a Dynes-like DoS with broadening parameter $\tilde{\gamma}_1/(2^{1/3}\sqrt{3})$, but we stress that the functional form of this broadening is different from the one produced by a Dynes parameter, since a hard gap persists. Next order corrections to the gap, DoS maximum value, and its position are discussed in Appendix~\ref{app:higherOrderExpansion}.

Figure~\ref{fig:Figure3}(a) displays the DoS of the two films in the close proximity of the gap for the same parameters used in Fig.~\ref{fig:Figure2}. The solid curves give the DoS obtained solving numerically the Usadel equations. For $\mathrm{S}_1$ we also display some analytical approximations: both the solution of the cubic Eq.~\eqref{eq:depressedCubicWeak} (orange dotted), and the improved solution at the next order in perturbation theory (black dashed, see Appendix~\ref{app:expansionWeak}) approach 
the BCS-like solution with shifted gap $\tilde{\e}_g$ (dot-dashed), which gives a good approximation of $\mathrm{S}_1$ over an extended energy range [cf. Fig.~\ref{fig:Figure2}(a)]. 
\begin{figure}
    \centering
\includegraphics{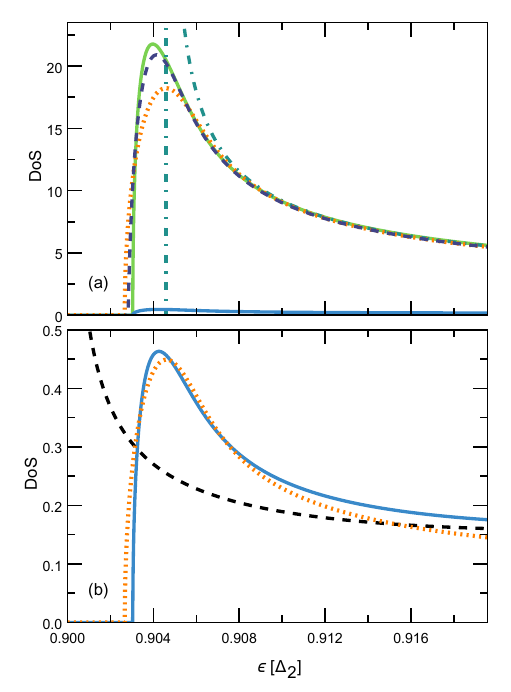}
    \caption{DoS in (a) $\mathrm{S}_1$ and (b) $\mathrm{S}_2$ close to the gap $\e_g$. The solid curves, computed solving numerically Eq.~\eqref{eq:UsadelAngular}, are as in Fig.~\ref{fig:Figure2}, since the parameters are the same; different vertical scales are used in the two panels for better readability. In panel (a), the dot-dashed curve gives the approximate BCS-like DoS for $\mathrm{S}_1$ with shifted gap $\tilde{\e}_g$ [Eq.~\eqref{eq:tildeEpsWeak}], already plotted in Fig.~\ref{fig:Figure2}.
    The orange dotted curve is the analytical approximations obtained from the solution of Eq.~\eqref{eq:depressedCubicWeak}, while the black dashed one is calculated at the next order in perturbation theory [see Eq.~\eqref{eq:DoShigherOrder} in Appendix~\ref{app:expansionWeak}].
    In panel (b), the orange dotted curve is computed by substituting $\mathcal{N}_1$ from the the solution to Eq.~\eqref{eq:depressedCubicWeak} into Eq.~\eqref{eq:relationDoSapp}. The black dashed curve is the analytical approximation computed using Eq.~\eqref{eq:explicitGamma} in Eq.~\eqref{eq:DynesAnsatz}, and is also shown in Fig.~\ref{fig:Figure2}. 
    }
    \label{fig:Figure3}
\end{figure}

\subsection{DoS and pair correlation function in the high-gap film close to the gap}
\label{sec:highGapDoSatEg}
We can use the results of Sec.~\ref{sec:lowerGapDoS} to obtain a more accurate expression for the DoS of the high-gap film close to the gap, where the treatment of Sec.~\ref{sec:smearingDoSX2} is not accurate. In particular, we are interested in the DoS for energies $\epsilon\gtrsim \epsilon_g$, where the DoS in the low-gap film is large, and so $|X_1|\gg 1$. We can relate the two DoSs using Eq.~\eqref{eq:XjVsXjtilde} with $j=2$; in particular, assuming $\tau_2^2(\dTwo^2-\e^2)\gg 1$ we can drop the first term under the square root in the denominator of Eq.~\eqref{eq:XjVsXjtilde}. Moreover, we approximate $\sqrt{X_1^2+1}\approx X_1\gg 1$ in it, and write 
\begin{align}
X_2&\simeq\frac{\tau_2 \e + X_1}{\tau_2\sqrt{\dTwo^2-\e^2}\sqrt{1+2X_1/[\tau_2(\e+\dTwo)]}}\nonumber
\\
&\simeq\frac{\tau_2 \e + X_1}{\tau_2\sqrt{\dTwo^2-\e^2}}\left[1-\frac{X_1}{\tau_2(\e+\dTwo)}\right]\,,
\label{eq:XjvsXjtildeApp}
\end{align}
where in the second line we expanded the last factor in the denominator for $|X_1|\ll \tau_2(\e+\dTwo)$, a condition which is asymptotically satisfied as the maximum DoS of $\mathrm{S}_1$ is proportional to $(\tau_1\dOne\tau_2\dTwo)^{1/3}$ (see Sec.~\ref{sec:lowerGapDoS}, where we assumed $(\tau_2\dTwo)^2 \gg \tau_1\dOne$). Taking the imaginary part of the second line in Eq.~\eqref{eq:XjvsXjtildeApp} we find at the leading order in $\tau_2\dTwo\gg 1$,
\begin{equation}
    \mathcal{N}_2(\e)\simeq \frac{\dTwo}{\dTwo+\e} \frac{\mathcal{N}_1(\e)}{\tau_2\sqrt{\dTwo^2-\e^2}}\, ,
    \label{eq:relationDoSapp}
\end{equation}
an expression accurate at this order in the range $\e_g < \e \lesssim \dOne+\sqrt{(\dTwo-\dOne)(\tilde\e_g-\dOne)/2}$ (the upper bound identifies the cross-over to the Dynes-like formula of Sec.~\ref{sec:smearingDoSX2}; see also Appendix~\ref{app:selfconsistence}. This result shows that for $\e\simeq \e_g$ the DoS in the high-gap film is small, $\mathcal{N}_2\ll 1$, even at the value where the DoS in the low-gap film has a maximum. We can use Eqs.~\eqref{eq:XjvsXjtildeApp} and \eqref{eq:relationDoSapp} to compute the pair correlation function $p_2=\Im[\sqrt{1+X_2^2}]$ in $\mathrm{S}_2$ near the gap $\e_g$; specifically, being $\mathcal{N}_2=\Im[X_2]\ll 1\leq\sqrt{1+\Re[X_2]^2}$, we have $p_2\simeq \Re[X_2]\mathcal{N}_2/\sqrt{1+\Re[X_2]^2}$. Keeping the leading order term in Eq.~\eqref{eq:XjvsXjtildeApp}, we find immediately
\begin{equation}
 p_2(\e)\simeq \e \mathcal{N}_2(\e)/\dTwo\simeq \frac{\e}{\dTwo+\e} \frac{\mathcal{N}_1(\e)}{\tau_2\sqrt{\dTwo^2-\e^2}}\, ,
\label{eq:relationp2DoS2}
\end{equation}
where in the second approximation we made use of Eq.~\eqref{eq:relationDoSapp}.

Figure~\ref{fig:Figure3}(b) shows the DoS of $\mathrm{S}_2$ in a different vertical scale compared to the panel (a) in the same figure. The numerically computed DoS (solid) is compared to the analytical approximations for energies close to the gap, namely Eq.~\eqref{eq:relationDoSapp} [using the solution of Eq.~\eqref{eq:depressedCubicWeak} to compute $\mathcal{N}_1$] (orange dotted) and the approximate solution obtained using the broadening of Eq.~\eqref{eq:explicitGamma} in the parametrization of Eq.~\eqref{eq:DynesAnsatz} (dashed black).

\section{Proximity effect in superconducting qubits with gap-asymmetric junctions
}
\label{sec:rates}

Having discussed in Sec.~\ref{sec:approxSolutions} the modification of the DoS in the two films due to the finite transparency of the barrier, 
here we show how the proximity effect can regularize divergencies in observables, taking as a concrete case quasiparticle tunneling~\cite{CatelaniPRB84,CatelaniSciPostReview} in gap-asymmetric transmon qubits~\cite{transmonPhysRevA76,DiamondPRXQuantum3,connolly2023coexistence,Krause24,nho2025}. Specifically, we address the regularizations of the divergence in the quasiparticle relaxation rate for qubit frequency ($\omega_{01}$) resonant to gap asymmetry~\cite{MarchegianiPRXQuantum} and of the related jump discontinuity in the qubit frequency shift~\cite{antonenko2025}.

\begin{figure}
    \centering
\includegraphics[width=0.46\textwidth]{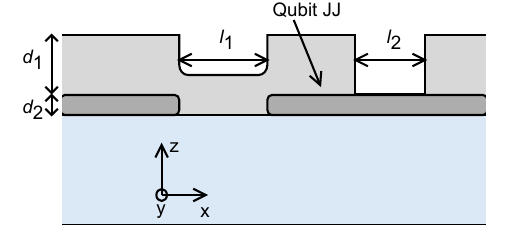}
    \caption{Schematic lateral view of a Josephson junction deposited with a double-angle evaporation technique. The junction is fabricated depositing a first superconducting film with thickness $d_2$ (dark gray) on top of an insulating substrate (light blue, typically silicon or sapphire). After an oxidation step to form the insulating barrier, a second film (light gray) is deposited on top of the first film with a different evaporation angle to form the junction. The thickness of the second film ($d_1$) is larger to ensure film continuity around the junction; this thickness mismatch determines a gap asymmetry for aluminum devices (see text). Due to the deposition technique, the two films form a bilayer everywhere, except for the two narrow regions of lengths $l_1$ and $l_2$ flanking the qubit junction.}
    \label{fig:Figure4}
\end{figure}

In Fig.~\ref{fig:Figure4} we depict the region around a Josephson junction in a typical superconducting qubit. The two films (represented by different shades of gray) are deposited consecutively, with an intermediate oxidation step to form the tunnel barrier of the Josephson junction. To ensure continuity of the film near the junction, the top film is generally thicker than the bottom layer. For the vast majority of qubits, the junction films are made out of aluminum, whose gap is strongly dependent on the thickness in the thin-film regime, roughly below 100~nm~\cite{CherneyThinAl,ChubovThinAl,MeserveyThinAl,Court_2007}. Then the thickness mismatch results in a gap asymmetry between the two superconductors. 

We now argue that the results obtained for a uniform bilayer can be used to analyze the impact of proximity effects in structures of the type shown in Fig.~\ref{fig:Figure4}. As already mentioned in Sec.~\ref{sec:Model}, neglecting the spatial dependence in the direction orthogonal to the junction plane [$z$-axis in Fig.~\ref{fig:Figure4}] is justified for thin films with thickness $d_i$ much smaller than the superconducting coherence length $\xi_i$,  $d_i\ll \xi_i$. In the geometry considered, the low-energy qubit modes are associated with current mostly flowing along the $x$-direction, and so variations in the $y$-direction can be disregarded. Finally, due to the double-angle deposition process, the two films form a bilayer almost everywhere, except for the two regions (with lengths $l_1$ and $l_2$) adjacent to the qubit Josephson junction (see arrow in Fig.~\ref{fig:Figure4}). When $l_1,\,l_2$ are smaller than or comparable to the coherence length, it is justified to neglect the spatial dependence in the $x$-direction as a first approximation. To go beyond this approximation or for longer $l_i$, the findings for the uniform bilayer can be used as a starting point to describe the spatial dependence along $x$, see for instance the discussion in Ref.~\cite{HosseinkhaniPRB97} for a normal-metal trap. Finally, we remark that in Sec.~\ref{sec:Model} we assumed no phase bias between the two films. When a current flows through a Josephson junction, a phase difference develops across it; however, since the area of the qubit junction is small compared to the total overlap area between the two films on either side of the junction, our modeling approximately captures the overall spectral properties of the structure.

\subsection{Relaxation rate}
In a single-junction transmon the relaxation rate at the leading order in $E_J\gg E_C$ can be expressed as~\cite{CatelaniPRB84,MarchegianiPRXQuantum} 
\begin{equation}
\Gamma_{10}=s_{10}^2(E_J/E_C)[S_1(\omega_{01})+S_2(\omega_{01})]\,,
\label{eq:Gamma10}
\end{equation}
where the dependence of the matrix element $s_{10}\simeq (E_C/8E_J)^{1/4}$ on the qubit parameters, such as the Josephson ($E_J$) and charging ($E_C$) energies, is unaffected by the proximity effect (although $E_J$ itself is renormalized by it, see Appendix~\ref{app:JosEnergyRenormalization}). The spectral densities read:
\begin{align}
S_j(\w)=\frac{4g_T}{e^2}\int_{\rm {max}(\e_g,\e_g-\w)}^{+\infty}\!\!\!\!\!d\e & \,\mathcal{N}_j(\e)\mathcal{N}_{\bar{j}}(\e+\w)F_{j\bar{j}}(\e,\e+\w)\nonumber\\
&\times f_j(\e)[1-f_{\bar j}(\e+\w)]\, ,
\label{eq:Sqp}
\end{align}
where $g_T$ is the normal state conductance of the junction, $e$ is the absolute value of the electron charge, $j={1,2}$ (with $\bar{1}=2$ and vice versa) denotes the tunneling quasiparticle's initial location being superconductor $\mathrm{S}_j$, and $f_j(\e)$ is the quasiparticle distribution function in $\mathrm{S}_j$.
In contrast to the BCS case in which proximity is ignored (see for instance Ref.~\cite{MarchegianiCommunPhys}), in the lower limit of the integral the common gap $\e_g$ replaces the order parameters $\Delta_1,\Delta_2$, and the coherence factors $F_{j\bar{j}}$ are generally expressed as~\cite{PhysRevB85-Quasiparticle-NonBCS} 
\begin{equation}
F_{j\bar{j}}(\e,\e')=\frac{1}{2}\left[1+\frac{p_j(\e)p_{\bar{j}}(\e')}{\mathcal{N}_j(\e)\mathcal{N}_{\bar{j}}(\e')}\right]\, .
\label{generalizedBCSfactors}
\end{equation} 
Ignoring the proximity effect and introducing $\wLR = \dTwo-\dOne$, the rate $\Gamma_{10}$, once normalized by the thermal quasiparticle density in the low-gap electrode $x_\mathrm{qp}^\mathrm{th} =\sqrt{2\pi T/\dOne}\  e^{-\dOne/T}$, is exponentially suppressed with $(\wLR-\w_{01})/T$ for $\wLR>\omega_{01}$, and diverges [through $S_1(\omega)$] when these two frequencies are resonant~\cite{MarchegianiCommunPhys}. Measurements of parity-switching and relaxation rates in transmons with gap-asymmetric junctions show a finite lifetime at resonance~\cite{DiamondPRXQuantum3,McEwenPRL133}; in what follows we show how proximity regularizes the divergence in $S_1(\omega)$. Near the resonant condition $\w=\wLR$, the spectral density can be approximated as 
\begin{align}
&S_1(\w)\simeq\frac{2 g_T \sqrt{\dOne\dTwo}
}{e^2}e^{-\tilde\e_g/T} 
\Re
\left\{
e^{\frac{z_0(\omega)}{2T}}
K_0\left[\frac{z_0(\w)}{2T}\right]
\right\},  
\label{eq:S_1PlusRelaxationExplicit}
\end{align}
where 
$z_0(\w)=\omega+\tilde{\e}_g-\dTwo+i\gamma_{2,0}$ and $K_0[z]$ is the modified Bessel function of the second kind (see Appendix~\ref{app:detailsApprox} for details on the derivation). Note that ignoring proximity ($\tau_i \to \infty$), using that $\Re[K_0[x]]=K_0[|x|]$ for a real argument $x$~\cite{abramowitz1968handbook}, this expression reproduces the one reported in Ref.~\cite{MarchegianiCommunPhys} (where we further approximated $\sqrt{\dOne\dTwo}\simeq\bar\Delta=(\dOne+\dTwo)/2$, assuming $\dTwo-\dOne\ll\bar\Delta$),
\begin{align}
S_{1,\rm BCS}(\w)\simeq\frac{2g_T\sqrt{\dOne\dTwo}}{e^2}e^{-\frac{\dOne+\dTwo-\w}{2T}}
K_0\left[\frac{|\w-\wLR|}{2T}\right]. 
\label{eq:S_1PlusBCS}
\end{align}
Close to its maximum, occurring for $\omega=\Delta_2-\tilde{\e}_g$, Eq.~\eqref{eq:S_1PlusRelaxationExplicit} approximately reads:
\begin{align}
S_1(\w)&\simeq-\frac{2g_T\sqrt{\dOne\dTwo}}{e^2}e^{-\tilde\epsilon_g/T}
\nonumber\\
&\times
\ln\left[\frac{e^{\gamma_E}}{4T}\sqrt{(\w-\dTwo+\tilde\e_g)^2+\gamma_{2,0}^2}
\right],
\label{eq:regularizedDiv}
\end{align}
regularizing the divergence occurring in Eq.~\eqref{eq:S_1PlusBCS} for BCS DoSs
(the numerical coefficients inside the logarithm are accurate for $\gamma_{2,0} \gg \tilde\gamma_1$).

\begin{figure}
    \centering
\includegraphics{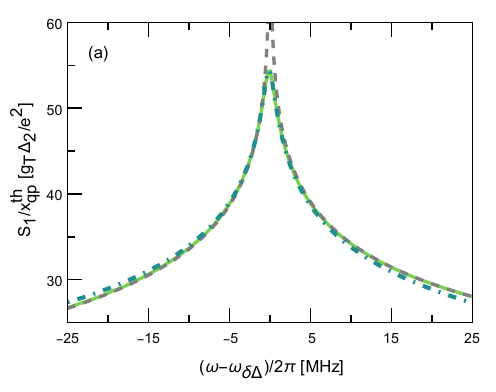}
    \caption{Regularization of the divergence in the qubit relaxation rate. Quasiparticle spectral density function $S_1$ (divided by the thermal quasiparticle density) vs frequency for a gap-asymmetric transmon. The result of numerical computation of the rate obtained by solving the Usadel equation (solid light green) is compared with the analytical approximation ignoring the proximity effect Eq.~\eqref{eq:S_1PlusBCS} (dashed gray), and the leading order approximation including proximity Eq.~\eqref{eq:regularizedDiv} (dot-dashed dark green). Parameters: $\dTwo/h=50~$GHz, $\dOne=0.9\dTwo$, $\tau_1=10^5/\dTwo$, $\tau_2=2\times10^4/\dTwo$, $T=0.01~\dTwo$.
    }
    \label{fig:Figure5}
\end{figure}

Figure~\ref{fig:Figure5} displays the frequency dependence of the spectral density function around the resonant condition $\omega=\omega_{\delta\Delta}\simeq 0.1\dTwo$. For visualization purposes, we divide this quantity by the thermal quasiparticle density $x_\mathrm{qp}^\mathrm{th}$ in $\mathrm{S}_1$. 
The solid curve in the plot has been obtained by using Eq.~\eqref{eq:Sqp}, computing the DoS and the pair correlation functions from the numerical solution of the Usadel equation Eq.~\eqref{eq:usadelXj}. The dashed curve gives the analytical approximation in Eq.~\eqref{eq:S_1PlusBCS} that ignores the proximity effect, $\tau_i \to\infty$. The formula accurately captures the behavior except close to $\wLR$; in this region, our approximation in Eq.~\eqref{eq:regularizedDiv} (dot-dashed) captures the cutoff of the diverging rate due to the proximity effect. 
\subsection{Frequency shift}
We now move to the regularization of the jump discontinuity in the qubit frequency shift. The spectral densities of Eq.~\eqref{eq:Gamma10} are related to the real part of the classical admittance of a Josephson junction~\cite{CatelaniPRB84},  
\begin{equation}
\Re[Y_{\rm qp}(\omega)]=\frac{e^2}{2\omega}[S(\omega)-S(-\omega)]\, ,
\label{eq:ReYqp}
\end{equation}
which characterizes the dissipative response (due to quasiparticles) of a Josephson junction to an AC drive [we defined $S(\omega)=S_1(\omega)+S_2(\omega)$]. Quasiparticles also modify the non-dissipative response of the junction, affecting the transition frequencies of superconducting qubits~\cite{CatelaniPRB84}; when the quasiparticle density is small ($f_j\ll 1$), it can be shown that the frequency shift for the transition between ground and first excited state can be expressed as~\cite{CatelaniPRB84,kurilovich2025correlatederrorburstsgapengineered}
\begin{equation}
\delta\omega=-\frac{1}{2C}{\Im}[Y_{\rm qp}(\omega_{01})]\,,
\label{eq:frequencyShift}
\end{equation}
where $C=e^2/2E_C$ is the transmon capacitance and $\Im[Y_{\rm qp}]$ is the quasiparticle contribution to the imaginary part of the junction admittance. This contribution includes a purely inductive component ($\Im[Y_{\rm qp}^{\rm ind}]\propto 1/\omega$) and a dynamic component ($\Im[Y_{\rm qp}^{\rm dyn}]$) which vanishes at $\omega=0$. For our goals, we can disregard $\Im[Y_{\rm qp}^{\rm ind}]$, which is regular for $\omega=\omega_{\delta\Delta}$. The dynamic component can be computed using the Kramers-Kronig relation
\begin{equation}
\Im[Y_{\rm qp}^{\rm dyn}]=-\mathcal{H}[\Re[Y_{\rm qp}(\omega^\prime)]](\omega)=\mathcal{P}\frac{1}{\pi}\int_{-\infty}^{\infty}\!\!d\omega^\prime \frac{\Re[Y_{\rm qp}(\omega^\prime)]}{\omega^\prime-\omega},
\label{eq:ImYqpintegral}
\end{equation}
in which $\mathcal{H}[Y(\omega')][\omega]$ denotes the Hilbert transform of the function $Y(\omega)$~\cite{bookHilbertTransform} and $\mathcal{P}$ the Cauchy principal value of the integral. When proximity effects are neglected, the dynamic component for $\omega>0$ can be approximated as~\cite{antonenko2025}
\begin{align}
&\Im[{Y_{\rm qp,BCS}^{\rm dyn}}]=\pi g_T
\frac{\bar\Delta}{\omega}e^{\frac{-\bar\Delta}{T}}\bigg\{e^{-\frac{\w}{2T}}I_0\left[\frac{\w+\omega_{\delta\Delta}}{2T}\right] \nonumber \\ &\qquad -2I_0\left[\frac{\omega_{\delta\Delta}}{2T}\right]
+e^{-\frac{|\omega-\omega_{\delta\Delta}|}{2T}}\left[\cosh\frac{\omega_{\delta\Delta}}{2T}-h(\omega-\omega_{\delta\Delta}) 
\right. \nonumber \\ & \qquad \times \left.
\sinh\frac{\omega_{\delta\Delta}}{2T}\right] 
I_0\left[\frac{|\w-\omega_{\delta\Delta}|}{2T}\right]\bigg\},
\label{eq:ImYqpBCS}
\end{align}
where $h(x)={\rm sign}(x)$, and $I_0[z]$ is the modified Bessel function of the first kind. Equation~\eqref{eq:ImYqpBCS} is an expression valid for $\omega_{\delta\Delta},T\ll\dOne,\dTwo$, and is obtained by combining Eqs.~\eqref{eq:ReYqp} and \eqref{eq:ImYqpintegral} and using the BCS spectral functions $S_{j,{\rm BCS}}$ in Eq.~\eqref{eq:ReYqp}. Within our approximations, $S_{1,\rm BCS}$ is given by Eq.~\eqref{eq:S_1PlusBCS} while $S_{2,\rm BCS}(\omega)$ can be obtained replacing $\omega_{\delta\Delta}\to-\omega_{\delta\Delta}$ in Eq.~\eqref{eq:S_1PlusBCS} [see, for instance, Ref.~\cite{MarchegianiCommunPhys}], in both cases approximating $\sqrt{\dOne\dTwo}\simeq\bar\Delta$ in the prefactor consistent with the assumption $\omega_{\delta\Delta}\ll \dOne,\dTwo$. We note that Eq.~\eqref{eq:ImYqpBCS} reduces to Eq.~(27) of Ref.~\cite{CatelaniPRB84} for $\omega_{\delta\Delta}=0$. 
Note that Eq.~\eqref{eq:ImYqpBCS} has a jump discontinuity for $\w=\wLR$ (due to the sign function $h$); when proximity effects are included, this term is approximately replaced by
\begin{equation}
h_{\rm prox}(\omega,\gamma_{2,0})=\frac{2}{\pi}\arctan\left(\frac{\omega}{\gamma_{2,0}}\right)
\label{eq:jprox}
\end{equation}
which is continuous for $\w=\wLR$ (details are provided in Appendix~\ref{app:detailsApprox}).
\begin{figure}
    \centering
\includegraphics[width=\columnwidth]{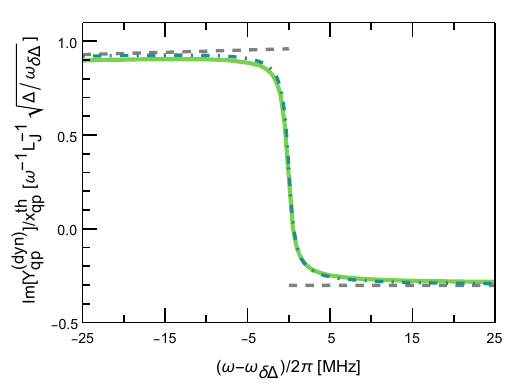}
    \caption{Regularization of the jump-discontinuity in the qubit frequency shift. Dynamic component of the imaginary part of the classical admittance of the Josephson junction (divided by the thermal quasiparticle density) vs frequency for a gap-asymmetric transmon. The result of numerical computation of the rate obtained by solving the Usadel equation (solid light green) is compared with the analytical approximation ignoring the proximity effect Eq.~\eqref{eq:ImYqpBCS} (dashed gray), and the leading order approximation including proximity [replacing $h\to h_{\rm prox}$, see Eq.~\eqref{eq:jprox}, in Eq.~\eqref{eq:ImYqpBCS}] (dot-dashed dark green). In the y-axis normalization, $L_J=[2 g_T\dOne\dTwo K(|\dOne-\dTwo|/2\bar{\Delta})/\bar{\Delta}]^{-1}$, where $K(x)$ is the complete elliptic integral of the first kind, is the Josephson inductance in the absence of quasiparticles~\cite{antonenko2025}, which is well approximated by $L_J\simeq(\pi g_T\bar\Delta)^{-1}$~\cite{CatelaniPRB84} in the limit $|\dOne-\dTwo|\ll \bar{\Delta}$ here considered. Parameters as in Fig.~\ref{fig:Figure5}.
    }
    \label{fig:Figure6}
\end{figure}

Figure~\ref{fig:Figure6} displays the dynamic component of the imaginary part of the junction admittance around the resonant condition $\omega=\omega_{\delta\Delta}\simeq 0.1\dTwo$. As for the spectral density in Fig.~\ref{fig:Figure5}, we divide this quantity by the thermal quasiparticle density $x_\mathrm{qp}^\mathrm{th}$ in $\mathrm{S}_1$ for visualization purposes. The solid curve in the plot has been obtained by using Eqs.~\eqref{eq:ImYqpintegral}, \eqref{eq:ReYqp}, and \eqref{eq:Sqp}, computing the DoS and the pair correlation functions from the numerical solution to the Usadel equation Eq.~\eqref{eq:usadelXj}. The dashed curve gives the analytical approximation in Eq.~\eqref{eq:ImYqpBCS} that ignores the proximity effect, $\tau_i \to\infty$; the formula captures the behavior of $\Im[Y_{\rm qp}^{\rm dyn}]$ (with deviations at the percent level) except close to $\wLR$; in this region, our approximation obtained replacing $h\to h_{\rm prox}$ in Eq.~\eqref{eq:ImYqpBCS} (dot-dashed) captures the regularization of the jump-discontinuity due to the proximity effect. The percentage-level deviations are due to subleading terms in $\omega_{\delta\Delta}/\bar\Delta$ which were neglected in the spectral densities $S_{j,\rm BCS}$ (the subleading contributions are reported in the Supplementary of Ref.~\cite{MarchegianiCommunPhys}). In closing this section, we remark that, while $\Im[Y_{\rm qp}^{\rm dyn}]$ can takes both positive and negative values, $\Im[Y_{\rm qp}]=\Im[Y_{\rm qp}^{\rm dyn}]+\Im[Y_{\rm qp}^{\rm ind}]$ is always positive, so that the qubit frequency shift is negative [cf. Eq.~\eqref{eq:frequencyShift}].

\section{Conclusions}
\label{sec:Conclusions}

In this work, we theoretically investigated proximity effects in bilayers composed of two thin superconducting films with asymmetric gaps. Using a quasi-classical description of the bilayer, in Sec.~\ref{sec:approxSolutions} we derived approximate solutions to the Usadel equations in the weak-coupling regime, as suitable, for instance, for superconductors separated by a thin insulating barrier. Our approximations give an accurate description of the DoS and the pair correlation functions in the two films; in particular, our results capture the broadening of the BCS singularities in the two films. At the gap the square root-singularity is transformed into a threshold; the DoS gap is the same in both films and its energy is slightly higher [by a quantity proportional to the small parameter $(\tau_1\dOne)^{-1}$] than the low gap when proximity effects are neglected. The BCS singularity in the high-gap film is broadened in a way formally equivalent to a Dynes parameter (with an energy modulation which mainly affects energies close to the gap), generalizing previous findings for normal-superconducting bilayers~\cite{HosseinkhaniPRB97}. In both films, the broadening is monotonically increasing with the asymmetry of the order parameters and vanishes in the gap-symmetric case. The broadening of the BCS singularities regularizes the logarithmic divergence in the convolution integral of the DoSs. As an example, in Sec.~\ref{sec:rates} we investigated the regularization of the quasiparticle-induced decay rate, and the related smoothing of the jump-discontinuity in the frequency shift, for transmons with frequency resonant to the asymmetry between the order parameters.     

Our modeling is expected to approximately describe a wide class of Josephson junction based devices, see the discussion at the beginning of Sec.~\ref{sec:rates}. Our results can find application in interpreting the current-voltage characteristic of superconducting junctions, and the performance of hybrid superconductors interferometers~\cite{Ligato_2017}. Our findings are also valuable for detection applications, since superconducting bilayers can be exploited to develop kinetic inductance detectors~\cite{Zhao_2018_KID,Wang_KID} and transition edge sensors~\cite{Lolli2016}; for the latter, our analysis should be extended to temperatures comparable to the superconducting order parameters. Generalization to multilayer structures would also be of interest for such applications~\cite{Cardani2018,Pesce2026}.

\acknowledgments   

This work was supported in part by the German Federal Ministry of Research, Technology and Space (BMFTR), funding program ``Quantum Technologies -- From Basic Research to Market,'' project QSolid (Grant No. 13N16149).

\section*{Data Availability}

The data used to generate figures 2, 3, and 5-8 are publicly available~\cite{zenodo}.

\appendix
\section{On the Dynes parametrization}
\label{app:Dynes}
In Sec.~\ref{sec:Dynes_param} we have introduced a Dynes-like parametrization for the Green's function in terms of which the Usadel equation takes the form of a self-consistency equation, Eq.~\eqref{eq:iterativeFormGammaj}, for the energy-dependent broadening $\gamma_j$. In Sec.~\ref{sec:smearingDoSX2}, we have derived the leading-order approximation for $\gamma_2(\epsilon)$ in the weak-coupling regime; here we discuss the validity of that approximation and how to improve on it.

\subsection{Energy range of validity for Eq.~(\ref{eq:explicitGamma})}
\label{app:energyRangeWeakDynes}
In Sec.~\ref{sec:smearingDoSX2}, we approximate $\mathcal{G}(\e;i\gamma_2)\simeq\tau_1^2(\dOne^2-\e^2)$ [valid under the condition of Eq.~\eqref{eq:conditionX2weak}], and extracting the square root with negative imaginary part, we immediately find Eq.~\eqref{eq:explicitGamma}.
The inequality leading to this expression doesn't hold for energy too close to the low-gap order parameter, since the left-hand side of Eq.~\eqref{eq:conditionX2weak} vanishes for $\e=\dOne$. Using Eq.~\eqref{eq:DynesAnsatz}, we rewrite the inequality in Eq.~\eqref{eq:conditionX2weak} as 
\begin{equation}
\tau_1^2 (\e^2-\dOne^2)\gg \left|1+2\tau_1\dOne\frac{1-\e(\e +i\gamma_2)/(\dOne\dTwo)}{\sqrt{1-(\e +i\gamma_2)^2/\dTwo^2}}\right|\,.
\label{eq:conditionX2weakDynes}
\end{equation}
To investigate the range of validity of the inequality, we introduce $\kappa= (\e-\dOne)/\dOne$, with $0<\kappa\ll 1$. The size of the second term in the right-hand side of Eq.~\eqref{eq:conditionX2weakDynes} in the vicinity of $\e=\dOne$ depends on the ratio in the square root term, which reads [using Eq.~\eqref{eq:explicitGamma}]
\begin{equation}
\frac{\e+i\gamma_2}{\dTwo}=\frac{i+\tau_2\dOne\sqrt{\kappa (\kappa+2)}}{i+\tau_2\dTwo\sqrt{\kappa (\kappa+2)}}(1+\kappa)\, .
\end{equation}
For energies such that $\kappa\gg 1/2(\tau_2\dOne)^2$, we obtain the estimate $\e+i\gamma_2\simeq \dOne$; substituting this expression into Eq.~\eqref{eq:conditionX2weakDynes}, and assuming $\sqrt{(1-\delta)/(1+\delta)}\gg 1/2\tau_1\dOne$, we obtain the inequality in Eq.~\eqref{eq:conditionGammaExplicit}. Replacing $\e\to\tilde\e _g$ [cf. Eq.~\eqref{eq:tildeEpsWeak}] in the lower bound on $\kappa$, we find the condition $\sqrt{(1-\delta)/(1+\delta)}\gg \tau_1  /2\tau_2^2\dOne$,
which can put an additional limitation on the minimal gap asymmetry. 
\subsection{Iterative approach}
\label{app:iterativeGamma}
\begin{figure}
    \centering
\includegraphics{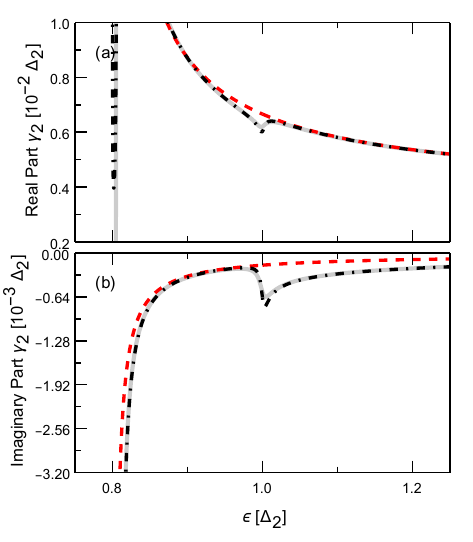}
    \caption{Real and imaginary part of $\gamma_2(\e)$ in the weak coupling regime. Solid curves, obtained through numerical solution of the Usadel equation [Eq.~\eqref{eq:usadelXj}], are compared with the analytical approximation in Eq.~\eqref{eq:explicitGamma} (dashed), and with the improved approximation replacing Eq.~\eqref{eq:explicitGamma} in the right-hand side of Eq.~\eqref{eq:iterativeFormGammaj} (dot-dashed). Parameters: $\dOne=0.8\dTwo$, $\tau_1=\tau_2=50/\dTwo$.
    }
    \label{fig:Figure7}
\end{figure}
The last term on the right-hand side of Eq.~\eqref{eq:squareRootGammaj} (with $j=2$) reaches its largest value close to $\e=\Delta_2$, where it approximately reads $2\tau_1\sqrt{\tau_2\dTwo(\dTwo^2-\dOne^2)}$, making the approximation in Eq.~\eqref{eq:conditionX2weak} slightly less accurate than for other energies [cf. Fig.~\ref{fig:Figure2}(b)]. Due to its structure, Eq.~\eqref{eq:iterativeFormGammaj} can be used to obtain improved approximations for $\gamma_2(\e)$ (and so for $X_2$) iteratively.
In Fig.~\ref{fig:Figure7} we display the real (panel a) and the imaginary (panel b) parts of $\gamma_2(\e)$: the solid curves are obtained solving numerically the Usadel equation [Eq.~\eqref{eq:usadelXj}] for $X_2$ and consequently computing $\gamma_2(\e)=-i[-\e+\dTwo X_2(\e)/\sqrt{1+X_2^2(\e)}]$ [see Eq.~\eqref{eq:DynesAnsatz}]; the dashed and dot-dashed approximations are given by Eq.~\eqref{eq:explicitGamma} and the expression obtained inserting Eq.~\eqref{eq:explicitGamma} in the right-hand side of Eq.~\eqref{eq:iterativeFormGammaj}, respectively. 

An alternative approach for the iterative solution of the Usadel equation Eq.~\eqref{eq:iterativeFormGammaj} in the weak-coupling regime $\tau_j\Delta_j\gg 1$ consists in setting $\gamma_2=0$ in the right-hand side of Eq.~\eqref{eq:iterativeFormGammaj} as the starting point for the iterative procedure, reading explicitly
\begin{equation}
\gamma_2^{(0)}=-i\frac{\tau_{1}\e(\Delta_2-\Delta_{1})}{\tau_{1}\Delta_{1}+\tau_2\Delta_2\sqrt{\mathcal{G}_2(\e;0)}}\,.
\end{equation}
This expression yields a good approximation at all energies, except in proximity of $\e=\Delta_2$ and close to the gap. Specifically, at $\e=\Delta_2$ the iterative procedure cannot converge uniformly to the numerical solution, since it returns $\gamma_2^{(n)}=0$ at every iteration. Moreover, as $\e$ approaches the gap the zeroth iteration gives a real-valued $\gamma_2$, effectively resulting in a singular DoS at the (approximate) gap energy $\e\simeq \tilde\e_g$ at which $\mathcal{G}_2(\e;0)$ changes sign.

\section{Weak-coupling regime: asymptotic calculations of the gap and approximations to the density of states.}
\label{app:expansionWeak}
In this Appendix, we detail the perturbative approach used to derive the analytical approximations presented in the main text. As discussed in Sec.~\ref{sec:lowerGapDoS}, to determine the spectral properties of $\mathrm{S}_1$, including an accurate approximation for the size of the gap, we consider the variable 
\begin{equation}
u=\sqrt{X_1^2+1}-X_1 \, ,
\label{eq:uDef}
\end{equation}
so to express in a ``symmetrized'' form the quantities
\begin{align}
X_1&=\frac{1}{2u}(1-u^2)\, ,
\label{eq:X1vsu}\\
\sqrt{1+X_1^2}&=\frac{1}{2u}(1 + u^2)\, ,
\label{eq:X1Sqvsu}
\end{align}
in Eq.~\eqref{eq:selfXj}, which are related to the normal and anomalous Green's function (see Sec.~\ref{sec:Model}). Clearly, $u$ is real (and positive), when $X_1$ has zero imaginary part; in other words, the gap is identified by the change in the nature of the solution to the Usadel equation from purely real to complex, both in terms of $X_1$ or $u$.
For uncoupled superconductors ($\tau_j\Delta_j\to +\infty$, corresponding to the BCS case), we have
\begin{equation}
u_{\rm BCS}=\sqrt{\frac{\dOne-\epsilon}{\dOne+\epsilon}}\, ,
\label{eq:uBCS}
\end{equation}
which is zero at the gap edge ($\e=\dOne$); thus, in the weak-coupling limit, we expect $|u|\ll 1$ for energies close to the gap (notably, this inequality holds also in the strong-coupling regime, cf. Appendix~\ref{app:strong}). 

Substituting Eqs.~\eqref{eq:X1vsu} and~\eqref{eq:X1Sqvsu} in Eq.~\eqref{eq:selfXj}, after squaring we write the Usadel equation in the form
\begin{align}
&\beta_1^2[1-\tilde\epsilon-u^2(1+\tilde\epsilon)]^2[u(1+\beta_2^2(1-\delta^2\tilde\epsilon^2))+\beta_2(1-\delta\tilde\epsilon)
\nonumber\\
&+\beta_2(1+\delta\tilde\epsilon)u^2]
=\beta_2^2[u^2(1+\delta\tilde\epsilon)-(1-\delta\tilde\epsilon)]^2u \, ,
\label{eq:UsadelScaledU}
\end{align}
where
$\beta_j=\tau_j\Delta_j$, $\delta=\dOne/\dTwo$, and $\tilde\epsilon=\epsilon/\dOne$.

In the limit of weak coupling $\beta_j\gg 1$, we can approximate $1+\tilde\epsilon\approx 2$, $1+\delta\tilde\epsilon\approx1+\delta$ and $1-\delta\tilde\epsilon\approx 1-\delta$ for energies close to the gap; the last approximation is valid under the additional condition $\sqrt{1-\delta^2}/\delta \gg1/(\tau_1\dOne)$ [obtained requiring $\tilde\e-1\ll 1-\delta $, and replacing $\e\to \tilde{\e}_g$], so the two order parameters cannot be arbitrarily close. Rescaling the energy and the variable as 
\begin{equation}
    \varepsilon=\beta_1(\tilde\epsilon-1)\,,\quad \tilde u=\beta_1^{1/2}u \, ,
    \label{eq:rescalingOne}
\end{equation}
and defining $\tilde\beta_2=\beta_2(1+\delta)$ and $A=\sqrt{(1-\delta)/(1+\delta)}$, we can rewrite Eq.~\eqref{eq:UsadelScaledU} as 
\begin{align}
[2\tilde u^2+\varepsilon]^2 [\tilde u(1+A^2\tilde\beta_2^2 )+\frac{\tilde\beta_2}{\beta_1^{1/2}}(\beta_1A^2+ \tilde u^2)]&
\nonumber\\
=\tilde u \frac{\tilde\beta_2^2}{\beta_1^2}[\tilde u^2-A^2\beta_1]^2& \, .
\label{eq:UsadelScaledUapprox}
\end{align}
The solution to Eq.~\eqref{eq:UsadelScaledUapprox} at the leading order in $\beta_1,\tilde\beta_2\gg 1$ is obtained retaining only quadratic terms in $\beta_1\sim \tilde\beta_2$ in Eq.~\eqref{eq:UsadelScaledUapprox}, reading 
\begin{equation}
\label{eq:uLeadingWeak}
\tilde{u}=\frac{1}{\sqrt{2}}\sqrt{\sqrt{\frac{1-\delta}{1+\delta}}-\varepsilon}\,.
\end{equation}
This expression is equivalent to the uncoupled BCS solution [approximating $\e+\dOne\simeq2\dOne$, cf. Eq.~\eqref{eq:uBCS}] after the substitution $\dOne\to \dOne + \tilde\epsilon_g$, with $\tilde\epsilon_g$ given in Eq.~\eqref{eq:tildeEpsWeak}. In other words, we obtain the leading order correction to the gap discussed in Sec.~\ref{sec:lowerGapDoS}. The expression of Eq.~\eqref{eq:uLeadingWeak} is real for $\epsilon<\tilde{\epsilon}_g$ and purely imaginary for $\epsilon>\tilde{\epsilon}_g$, and still displays a square root divergence at the (approximate) gap value. As discussed in Sec.~\ref{sec:lowerGapDoS}, the divergence is regularized at the next-to-leading order. Since $\tilde{u}=0$ for $\epsilon=\tilde\epsilon_g$, a suitable variable change is necessary to perform a regular perturbation expansion, namely
\begin{equation}
\tilde{u} = \frac{t}{\beta^{1/6}}\,, \, \varepsilon-A = \frac{\eta}{\beta^{1/3}}\, ,
\label{eq:scalingNextToLeading}
\end{equation}
introducing $\beta=\tilde\beta_2^2/\beta_1$. The power law in the first equality in Eq.~\eqref{eq:scalingNextToLeading} has been determined using the method of dominant balance~\cite{bender2013advanced}.
Inserting Eq.~\eqref{eq:scalingNextToLeading} into Eq.~\eqref{eq:UsadelScaledUapprox}, the quadratic terms in $\beta_1$ vanish. Thus, at the leading order in $\beta\gg 1$, and using the assumption $\beta_1 A^2 \gg 1$, we obtain a depressed cubic equation (for later convenience, we rename $t\to t_0$ to denote the solution at this perturbative order)
\begin{equation}
\label{eq:depressedCubicWeakt}
t_0^3+\frac{\eta}{2} t_0 +\frac{A}{4}=0\,, 
\end{equation}
as discussed in Sec.~\ref{sec:lowerGapDoS} [see Eq.~\eqref{eq:depressedCubicWeak}]. The physically relevant solution is the root with negative imaginary part, which reads:
\begin{equation}
\label{eq:explicitRootCubic}
t_0=\frac{e^{-\frac{i\pi}{3}}}{2^{1/3}}\left(\frac{A}{4}+\sqrt{\frac{-\mathcal{D}}{27}}\right)^{\frac{1}{3}}-\frac{2^{1/3} \eta}{6 e^{-i\frac{\pi}{3}} }\left(\frac{A}{4}+\sqrt{\frac{-\mathcal{D}}{27}}\right)^{-\frac{1}{3}}\,,
\end{equation}
where $\mathcal{D}=-27 (A/4)^2-4(\eta/2)^3$ is the discriminant of the depressed cubic. Setting $\mathcal{D}=0$ we find the position of the gap
\begin{equation}
\eta_0=-\frac{3A^{2/3}}{2}\,.
\end{equation}
Expanding $t_0$ around $\eta_0$ we find $t_0\simeq A^{1/3}/2 -i\sqrt{\tilde\eta/6}$ where $\tilde\eta= \eta-\eta_0$; the DoS being proportional to $\Im[1/t_0]$, this gives rise to a square-root threshold. Note that Eq.~\eqref{eq:explicitRootCubic} asymptotically approaches (up to the factor $\beta^{1/6}$) Eq.~\eqref{eq:uLeadingWeak} -- an expression that returns a DoS of the BCS-like form (inverse square root) with shifted gap -- for energies sufficiently higher than the gap. More precisely, assuming $|\eta\,t_0|\gg A/2$, the solution to Eq.~\eqref{eq:depressedCubicWeakt} approximately reads $t_0\simeq-i\sqrt{\eta/2}$, which is manifestly equivalent to Eq.~\eqref{eq:uLeadingWeak} [cf. Eq.~\eqref{eq:scalingNextToLeading}]; the validity condition is $\eta\gg A^{2/3}/2^{1/3}$, or, more explicitly, $\e-\tilde\e_g\gg\{(1-\delta)/[2\beta(1+\delta)]\}^{1/3}/\tau_1$.

The perturbative solution in Eq.~\eqref{eq:explicitRootCubic} enables us to derive perturbative approximations for the DoS in the low-gap film.  Using Eqs.~\eqref{eq:rescalingOne} and \eqref{eq:scalingNextToLeading}, we have $u=t_0/\beta_1^{1/2}\beta^{1/6}$.  Since $|u|^2\ll 1$ close to the gap, we can approximate $X_1\approx 1/(2u)$ [cf. Eq.~\eqref{eq:X1vsu}], and the DoS reads $\mathcal{N}_1\simeq \beta_1^{1/2}\beta^{1/6}\Im[t_0^{-1}]/2$; note that the neglected $|u|^2$ term is of order $\beta_1^{-1}\beta^{-1/3}\ll 1$, and hence beyond accuracy of this asymptotic expansion. Expanding Eq.~\eqref{eq:explicitRootCubic} in powers of $\eta$, we can express the DoS as
 \begin{equation}
\mathcal{N}_1\simeq \frac{\sqrt{3}\beta_1^{1/2}\beta^{1/6}}{4^{2/3}A^{1/3}}\left(1-\sigma^2+\frac{2}{3}\sigma^3\right)\, ,
\label{eq:cubicDoS_S1}
 \end{equation}
 where we have defined $\sigma=2^{1/3}\eta/(3A^{2/3})$ (for $\eta=\eta_0$, we have $\sigma_0=-1/4^{1/3}$). The maximum of the DoS occurs at $\eta=0$, i.e., for $\epsilon=\tilde\epsilon_g$, and is proportional to $(\tau_1\dOne\tau_2\Delta_2)^{1/3}$, as discussed in the main text [cf. Eqs.~\eqref{eq:MaximumDosX1Weak} and \eqref{eq:minigapWeak}]. 

\subsection{Higher order expansion}
\label{app:higherOrderExpansion}
For the expansion at the following order, the equation to consider is a quintic, 
\begin{equation}\label{eq:quintic}
    \nu t^5 +(A+\nu\eta)t^3+\nu A t^2+\frac{\eta}{2}\!\left(A+\nu\frac{\eta}{2}\right)t+\frac{A}{2}\!\left(\nu\eta+\frac{A}{2}\right)=0,
\end{equation}
where $\nu=\beta^{-1/3}\ll 1$ (factors of $A$ could be removed from this equation by rescaling $t\to A^{1/3} t$, $\eta \to A^{2/3}\eta$, and $\nu \to A^{1/3} \nu$; the latter quantity then identifies the actual asymptotic parameter). By properly expanding around $\eta_0$, one can find the next order expressions for the gap position ($\eta_1$) and for $t$, namely:
\begin{equation}
\eta_1=\eta_0\left(1-\frac{\nu}{A^{1/3}}\right), \ t_1 = \frac{A^{1/3}}{2}\left(1-\frac{\nu}{A^{1/3}}\right) - i \sqrt{\frac{\eta-\eta_1}{6}},
\end{equation}
as one can check by substituting $\eta=\eta_1+\nu^2 \bar\eta$, $t=A^{1/3}/2+\nu\bar{t}$ into Eq.~\eqref{eq:quintic} end keeping terms of order $\nu^2$.
Restoring explicitly the dependence on the physical parameters, the position of the gap at this perturbative order reads
\begin{align}
&\frac{\epsilon_g-\tilde\epsilon_g}{\dOne} =-\frac{3}{2(\tau_1\dOne)^{2/3}(\tau_2\dTwo)^{2/3}}\frac{(1-\delta)^{1/3}}{1+\delta} \nonumber \\
&\times 
    \left[1-\frac{(\tau_1\dOne)^{1/3}}{{(\tau_2\dTwo)^{2/3}(1+\delta)^{1/2}(1-\delta)^{1/6}}}\right].
\end{align} 

Away from the gap, we can consider a regular expansion $t=t_0+\nu\tilde{t} +o(\nu)$, where $o(z)$ denotes the little $o$ symbol, i.e., $\lim_{\nu\to 0} o(\nu)/\nu=0$. Solving perturbatively the resulting equation, one finds 
\begin{equation}
\tilde{t}=-\frac{1}{4[1+4\eta t_0/(3A)]}\, .
\label{eq:t1solution}
\end{equation}
This formula is used to obtain the analytical approximation for the DoS displayed in Fig.~\ref{fig:Figure3}(a), namely
\begin{equation}
\mathcal{N}_1\simeq\frac{\beta_1^{1/2}\beta^{1/6}}{2}\Im\left[\frac{t_0-\nu \tilde{t}}{t_0^2}\right]
\label{eq:DoShigherOrder}
\end{equation}
where $t_0$ and $\tilde{t}$ are given in Eqs.~\eqref{eq:explicitRootCubic} and \eqref{eq:t1solution}, respectively. Using this expression
we can obtain the subleading correction to the position of the maximum of $\mathcal{N}_1$.
Since at the leading order the DoS is maximum at $\eta=0$, we consider a quadratic expansion of $t_0$ and $\tilde{t}$ in $\eta$, and we find for the DoS
\begin{equation}
\mathcal{N}_1\approx\frac{-4\eta^2 -12(2A)^{1/3}\nu\eta+9 [A\nu+(2A)^{4/3}]}{12\sqrt{3}\, 2^{2/3}A^{5/3}}\beta^{1/6}\beta_1^{1/2}\,.
\label{eq:quadraticEtaMaximum}
\end{equation}
The position of the maximum is immediately found by setting the first derivative of the numerator of Eq.~\eqref{eq:quadraticEtaMaximum} with respect to $\eta$  to zero, giving
\begin{equation}
\eta_{\rm max}=\frac{-3A^{1/3}\nu}{2^{2/3}}\, ,
\end{equation}
which can be rewritten explicitly in terms of the parameters $\tau_j,\Delta_j$ as
\begin{equation}
    \label{eq:posMaxS1}
    \epsilon_{\rm max}^{\mathrm{S}_1}=\tilde\epsilon_g-\frac{3\dOne^{2/3}}{2^{2/3}\tau_1^{1/3}\tau_2^{4/3}}\frac{(\dTwo-\dOne)^{1/6}}{(\dTwo+\dOne)^{3/2}}  \, .
\end{equation}
\section{Self-consistent gap calculation}
\label{app:selfconsistence}
Here we discuss the calculation of the self-consistent order parameters $\Delta_{j}$, $j=1,\,2$, [entering the Usadel equations Eq.~\eqref{eq:UsadelAngular}] in terms of their values $\Delta_{j,0}$ in the absence of proximity effect. For numerical calculations, we consider the self-consistency relation in the Matsubara representation; in practice, we replace $\epsilon\to i\omega_k$ and the integral with a summation over frequencies $\omega_k=2\pi T(k+1/2)$ in Eq.~\eqref{eq:selfGap}, obtaining
\begin{equation}
\frac{\Delta_j(T)}{\Delta_{j,0}}= \frac{\sum_{k= 0}\Re[\sin\theta_j(\omega_k)]}{\sum_{k=0} 
\frac{\Delta_{j,0}}{\sqrt{\Delta_{j,0}^2+\omega_k^2}}}\,, 
\label{eq:selfMatsubara}
\end{equation}
with $\Delta_{j,0}$ the BCS order parameter of superconductor $j$ at zero temperature in the absence of proximity effect. The solution of the two coupled equations in Eq.~\eqref{eq:selfMatsubara} is achieved iteratively: we solve the Usadel equations Eq.~\eqref{eq:UsadelAngular} for a initial guess $\{\Delta_{1}^{(0)},\Delta_{2}^{(0)}\}$, and generate a new pair of values for the order parameters using Eq.~\eqref{eq:selfMatsubara}, for instance after $n$ iterations we obtain $\{\Delta_{1}^{(n+1)},\Delta_{2}^{(n+1)}\}$. At each step, we evaluate $\max_{j=1,2}{|\Delta_{j}^{(n+1)}- \Delta_{j}^{(n)}|/\Delta_{j}^{(n)}}$ and we stop the procedure when this quantity is smaller than a certain threshold $\chi \ll 1$; we usually set $\chi=10^{-5}$ and evaluate the terms of the Matsubara sum up to $k=2000$.

\subsection{Analytical approximation at weak coupling}
For analytical calculations, 
we restrict our analysis to the low-temperature limit $T\ll \e_g$, where we can approximate ${\rm tanh}(\e/2T)\simeq {\rm sgn}(\e)$ in Eq.~\eqref{eq:selfGap}. In general, the two equations for $\dOne$ and $\dTwo$ [Eq.~\eqref{eq:selfGap} with $j=1$ and $2$] are coupled; nevertheless, we will see that, at the leading order in the asymptotic expansions for large $\beta_j=\tau_j\Delta_j$, $j=1,2$, the two equations are independent. We start with the low-gap film:we introduce an energy $\e^*$ such that $\min\{\dTwo-\dOne,\dOne\}\gg\e^*-\dOne\gg \tau_1^{-1}$.
For $\e \lesssim \e^*$ we approximate $p_1(\e)\simeq\mathcal{N}_1(\e)\simeq\beta_1^{1/2}\beta^{1/6}\Im[t_0^{-1}]/2$ (see Appendix~\ref{app:expansionWeak}). For $\e\gtrsim \e^*$, using Eq.~\eqref{eq:XjsqVsXjtilde} and the Dynes parametrization of $X_2(\e)$ [cf. Eq.~\eqref{eq:DynesAnsatz}], we write
\begin{align}
\sqrt{1+X_1^2} &= \frac{\tau_1\Delta_1 + \frac{\dTwo}{\sqrt{\dTwo^2-(\e+i\gamma_{2})^2}}}{\sqrt{1+\tau_1^2(\Delta_1^2-\e^2)+2\tau_1\frac{\Delta_1\Delta_2-\e (\e+i\gamma_{2})}{\sqrt{\dTwo^2-(\e+i\gamma_{2})^2}} }}\nonumber\\
&\simeq \frac{i\dOne}{\sqrt{\e^2-\dOne^2}}\left[1+\frac{\frac{\dTwo-\dOne}{\dOne}-\frac{\dOne\dTwo-\dOne^2-i\e\gamma_{2}}{\dOne^2-\e^2}}{\tau_1\sqrt{\Delta_2^2-(\e+i\gamma_{2})^2}}\right],
\label{eq:X1squaredApprox}
\end{align}
where in the second line we used $\tau_1^2(\e^2-\dOne^2)\gg 1$ (consistent with our assumptions on $\e^*$), and performed a linear expansion for $\tau_1\dOne\gg |[\Delta_2^2-(\e+i\gamma_{2})^2]^{-1/2}{\rm \max}\{\dTwo,2\dOne[\dOne\dTwo-\e(\e+i\gamma_{2})]/(\e^2-\dOne^2)\}|$ (which requires $\beta_1\gg\sqrt{\beta_2}$). Taking the imaginary part of Eq.~\eqref{eq:X1squaredApprox} and keeping the leading order correction in $\tau_1\dOne\gg 1$ [setting $\gamma_2=0$ in the equation], we find
\begin{equation}
p_1(\e)\simeq \frac{\dOne}{\sqrt{\e^2-\dOne^2}}\left[1+\frac{\e^2}{\e^2-\dOne^2}\frac{(\dTwo-\dOne)\theta(\dTwo-\e)}{\beta_1\sqrt{\Delta_2^2-\e^2}}\right].
\label{eq:p1Approx}
\end{equation}
As an aside, we note that in the limit $\gamma_2\to 0$ these expressions can also be obtained from the exact (at leading order) solution to Eq.~\eqref{eq:UsadelScaledU}, namely
\begin{equation}\label{eq:ub1exact}
u^2=\frac{1-\tilde\e+\beta_1^{-1}\sqrt{(1-\delta\tilde\e)/(1+\delta\tilde\e)}}{1+\tilde\e+\beta_1^{-1}\sqrt{(1+\delta\tilde\e)/(1-\delta\tilde\e)}}\,.
\end{equation}

With the above approximations, the integrals in the two regions can be expressed in a closed form, namely:
\begin{align}
\int_{\e_g}^{\e^*}p_1(\e)d\e&\simeq \frac{\beta_1^{1/2}\beta^{1/6}}{2}\int_{\e_g}^{\e^*}\Im(t_0^{-1})d\e
\label{eq:intp1belowestar}\\
&\simeq \frac{9\dOne A^{1/3}}{16\sqrt{3}\beta_1^{1/2}\beta^{1/6} }\mathcal{F}\left[\frac{2\tau_1\beta^{1/3}}{3A^{2/3}}
(\e^*-\tilde\e_g)\right]\,,
\nonumber
\end{align}
\begin{align}
\int_{\e^*}^{\omega_{D,1}}\!\!\!p_1(\e)d\e&\simeq \dOne\!\left[\mathrm{ArcCosh}\left(\frac{\omega_{D,1}}{\dOne}\right)\!-\mathrm{ArcCosh}\left(\frac{\e^*}{\dOne}\right)\!\right]\nonumber\\
&+\frac{\dTwo[F(\psi,\sqrt{1-\delta^2})-E(\psi,\sqrt{1-\delta^2})]}{\tau_1(\dOne+\dTwo)}\nonumber\\
&+\frac{\dOne^2}{\tau_1(\dOne+\dTwo)\e^*}\sqrt{\frac{\dTwo^2-\e^{*2}}{\e^{*2}-\dOne^2}}\,,
\label{eq:intp1aboveestar}
\end{align}
where 
\begin{align}
\mathcal{F}(x)&=x^4(1+\sqrt{1+x^3})^{-4/3}+4x(1+\sqrt{1+x^3})^{-1/3}
\nonumber\\
&+[3-\sqrt{1+x^3}](1+\sqrt{1+x^3})^{1/3}\,,
\label{eq:integrationImInvt0}
\end{align}
$\psi=\arcsin[\sqrt{(\dTwo^2-\e^{*2})/(\dTwo^2-\dOne^2)}]$, and $F(\theta,z)$ and $E(\theta,z)$ are the elliptic integral of the first and second kind, respectively~\cite{PrudnikovBook}. Note that, for $x\gg 1$, we have $\mathcal{F}(x)\simeq 16\sqrt{x}/3$, and the integral of Eq.~\eqref{eq:intp1belowestar} approximately reads $\sqrt{2\dOne(\e^*-\tilde\e_g)}$, which coincides with the integral of the simplified form of BCS DoS with renormalized gap between $\tilde\e_g$ and $\e^*$. Summing the two contributions of Eqs.~\eqref{eq:intp1belowestar} and \eqref{eq:intp1aboveestar}, terms depending explicitly on $\e^*$ cancel out at the leading order in $1/\beta_1\ll 1$. In fact, at this order in the perturbative expansion, the contribution in the second line of Eq.~\eqref{eq:intp1belowestar} is compensated by the sum of the last term in the square parentheses of the first line of Eq.~\eqref{eq:intp1aboveestar} and the one in third line of Eq.~\eqref{eq:intp1aboveestar}. Moreover, the elliptic integrals can be taken to be complete ($\psi=\pi/2$), ignoring subleading contributions. Finally, we can express the integral of $p_1(\e)$ as 
\begin{align} \label{eq:integrationp1}
& \int_{\e_g}^{\omega_{D,1}}\!\!\!\!d\e\,p_1 \simeq \dOne \mathrm{ArcCosh}\frac{\omega_{D,1}}{\dOne} \\ &  \qquad \qquad \ + \frac{K(\sqrt{1-\delta^2})-E(\sqrt{1-\delta^2})}{\tau_1(\delta+1)} \nonumber\\
&\simeq\dOne\left[\log\left(\frac{2\omega_{D,1}}{\Delta_1}\right) +\frac{K(\sqrt{1-\delta^2})\!-E(\sqrt{1-\delta^2})}{\beta_1(\delta+1)}\right],
\nonumber
\end{align}
where $K(z)=F(\pi/2,z)$, $E(z)=E(\pi/2,z)$, and in the approximation of the last line of Eq.~\eqref{eq:integrationp1} we assumed $\omega_{D,1}\gg\dOne$. Note that since $\dTwo$ appears only in the contribution suppressed by $\beta_1$, we can substitute there $\dTwo \to \Delta_{2,0}$. From Eq.~\eqref{eq:selfGap}, in this limit and without proximity we have $\Delta_{1,0}=2\omega_{D,1}\exp(-1/\nu_{0,1}\lambda_1)$, and using Eq.~\eqref{eq:integrationp1}, we find the leading order increase of the order parameter:
\begin{equation}
\dOne\simeq\Delta_{1,0}\left[1+\frac{K\left(\sqrt{1-\delta_0^2}\right)-E\left(\sqrt{1-\delta_0^2}\right)}{\tau_1 \Delta_{1,0}(1+\delta_0)}\right]
\label{eq:Delta1ApproxSCweak}
\end{equation}
where $\delta_0=\Delta_{1,0}/\Delta_{2,0}$. We note that the second term in the right hand side of this equation vanishes for $\Delta_{1,0}=\Delta_{2,0}$, as expected for gap-symmetric films.

Now we discuss the integral of the pair correlation function $p_2=\Im[\sqrt{X_2^2+1}]$ of the high-gap film: 
using the Dynes ansatz of Eq.~\eqref{eq:DynesAnsatz} (with $j=2$) we express the pair correlation function as 
\begin{equation}
p_2(\e)= \Im\frac{\Delta_2}{\sqrt{\Delta_2^2-[\e +i\gamma_2(\e)]^2}}\, ,
\label{eq:pair2Dynes}
\end{equation}
where $\gamma_2(\e)$ is well approximated by Eq.~\eqref{eq:explicitGamma_2nd} (except close to the gap,
see Sec.~\ref{sec:smearingDoSX2}). By symmetry, we can restrict $\e>0$ in the integral of Eq.~\eqref{eq:selfGap}; then, the lower limit can be set to the gap $\e_{g}$. To estimate the integral, we introduce the energies $\e^*, \, \e''$, satisfying the inequalities $ \tau^{-1}_1\ll\e^*-\dOne\ll\min\{\dOne,\dTwo-\dOne\}$ and $\gamma_{2,0}\ll\dTwo-\e''\ll\dTwo-\dOne$. Correspondingly, 
we split the integration range in three regions: (i) $\e_g<\e\lesssim\e^*$,
where we use Eq.~\eqref{eq:relationp2DoS2} and approximate $p_2(\e)\simeq p_2^{\rm (i)}(\e)=\e\mathcal{N}_1(\e)/[\tau_2(\dTwo+\e)\sqrt{\dTwo^2-\e^2}] $;
(ii) $\e^*\lesssim\e\lesssim\e''$, where $\gamma_2(\e)\ll (\Delta_2^2-\e^2)/\e$ and we can use a leading-order expansion $p_2(\e)\simeq  p_2^{\rm (ii)}(\e)=\e\Re[\gamma_2(\e)]\Delta_2/(\Delta_2^2-\e^2)^{3/2}$, with $\gamma_2$ in Eq.~\eqref{eq:explicitGamma_2nd}   [assuming $\beta_2 \gg (\beta_1)^{1/2}$]; (iii) $\e\gtrsim\e''$, where we can approximately neglect the energy dependence of $\gamma_2(\e)$, setting $\gamma_2(\e)\simeq \gamma_{2,0}$ in Eq.~\eqref{eq:pair2Dynes}. In region (i), we have $p_2^{(i)}(\e)\leq \e^*\mathcal{N}_1(\e)/[\tau_2(\dTwo+\dOne)\sqrt{\dTwo^2-\e^{*2}}]$. Since the integral of $\mathcal{N}_1$ is approximately $\sqrt{2\dOne(\e^*-\tilde{\e}_g)}$ [see Eqs.~\eqref{eq:intp1belowestar} and ~\eqref{eq:integrationImInvt0} and text that follows them], the contribution from this region is much smaller than $\dTwo/[\tau_2(\dOne+\dTwo)]$ and so subleading compared to the other contributions, as we show next. In the second region the integral gives
\begin{align}
\int_{\e^\prime}^{\e^{''}}\!\!d\e\, p_2^{\rm (ii)}= \frac{\dTwo}{\tau_2(\Delta_1+\Delta_2)}
\left[ 
\frac{\dTwo}{\e}\sqrt{\frac{\e^{2}-\Delta_1^2}{\Delta_2^2-\e^{2}}}
\right. \nonumber\\
 -E(\theta,\sqrt{1-\delta^2})
+\delta^2 F(\theta,\sqrt{1-\delta^2})
\Bigg]\Bigg{|}^{\e^{''}}_{\e^{*}}\,,
\label{eq:intp2Range2}
\end{align}
where $\theta=\arcsin[\sqrt{\dTwo^2(\e^2-\dOne^2)/\e^2(\dTwo^2-\dOne^2)}]$ is the angle variable in the incomplete elliptic integrals. In fact, in evaluting the terms involving the elliptic integrals we can replace $\e''\to\dTwo$ ($\theta \to \pi/2$) and $\e^*\to\dOne$ ($\theta\to 0$), since the neglected contributions are subleading. The integration for $\e>\e''$ can be evaluated explicitly, returning 
\begin{align}
\int_{\e''}^{\omega_{D,2}}\!d\e\, p_2^{\rm (iii)}= \dTwo{\Re} \left[{\rm ArcCosh}\left(\frac{\e+i\gamma_{2,0}}{\Delta_2}\right)
\right]\Bigg|_{\e''}^{\omega_{D,2}}.
\label{eq:intp2Range3}
\end{align}
Summing Eqs.~\eqref{eq:intp2Range2} and ~\eqref{eq:intp2Range3}, terms depending explicitly on $\e''$ cancel out at the leading order in $1/\beta_2\ll1$, reading asymptotically $\gamma_{2,0}\sqrt{\dTwo/[2(\dTwo-\e'')]}$ [with plus sign from the first term in square brackets in  Eq.~\eqref{eq:intp2Range2} and minus sign in Eq.~\eqref{eq:intp2Range3}]; the first term in square brackets in Eq.~\eqref{eq:intp2Range2} gives a subleading contribution for $\e=\e^*$. As a result, using that $\omega_{D,2}\gg \Delta_2,\,\gamma_{2,0}$, the leading order correction to the integral of $p_2$ due to the proximity effect reads:
\begin{align}
\int_{\e_g}^{\omega_{D,2}}d\e \,p_2\simeq \dTwo\Bigg[ &\log\left(\frac{2\omega_{D,2}}{\Delta_2}\right)
\\
&-\frac{E(\sqrt{1-\delta^2})-\delta^2 K(\sqrt{1-\delta^2})}{\tau_2(\Delta_2+\Delta_1)}\Bigg].\nonumber
\end{align}
Noting that in the no-proximity case $\Delta_{2,0}=2\omega_{D,2}\exp(-1/\nu_{0,2}\lambda_2)$, we find immediately
\begin{equation}
\Delta_2\simeq \Delta_{2,0}\left[1-\frac{E(\sqrt{1-\delta_0^2})-\delta_0^2 K(\sqrt{1-\delta_0^2})}{\tau_2 \Delta_{2,0} (1 +\delta_{0})}\right].
\label{eq:Delta2ApproxSCweak}
\end{equation}
In the limit $\delta_0,\,\Delta_{1,0}\to 0$, one recovers the leading-order expression for the order parameter suppression in the superconducting layer of a NS bilayer~\cite{McMillanPR175,FominovPRB63,HosseinkhaniPRB97}.

We close this appendix estimating the value at which the approximations for the pair correlation of $S_2$ in region (i) and (ii) match, i.e., $p_2^{\rm (I)}(\e)=p_2^{\rm(II)}(\e)$. Equating the central expression in Eq.~\eqref{eq:relationp2DoS2} [with $\mathcal{N}_1\simeq \sqrt{\dOne/2(\e-\tilde\e_g)}$, assuming $\e-\dOne\gg 1/\tau_1$] and $p_2^{\rm(II)}$, we find $\sqrt{\dOne}\sqrt{\e^2-\Delta_1^2}(\Delta_2-\e)=\e\sqrt{2(\e-\tilde\e_g)}(\Delta_2-\Delta_1)$. Assuming 
$\e-\Delta_1\ll \Delta_1$, we obtain $\e^*\simeq \Delta_1+\sqrt{(\Delta_2-\Delta_1)(\tilde{\e}_g-\Delta_1)/2}$, a result which was quoted in Sec.~\ref{sec:highGapDoSatEg}.

\section{Renormalization of the Josephson energy}
\label{app:JosEnergyRenormalization}
The Josephson energy of a Josephson junction is proportional to its critical current $I_J$, $E_J= I_J/2e$~\cite{tinkham} ($\hbar=1$). The critical current (at $T=0$) can be expressed in terms of the pair correlation functions as a double integral~\cite{Barone}:
\begin{align}
   I_J&=\frac{2g_T}{e\pi} \int_{0}^\infty d\e \int_{0}^\infty d\e^\prime \frac{p_1(\e)p_2(\e^\prime)}{\e+\e^\prime}.
    \label{eq:criticalCurrent}
\end{align}
When proximity effects are ignored, we have $p_j=\Delta_{j,0}\theta(\e^2-\Delta_{j,0}^2)/\sqrt{\e^2-\Delta_{j,0}^2}$ and performing the double integral one gets~\cite{Barone}
\begin{equation}
I_{J,\rm BCS}=\frac{2g_T}{e}\frac{\Delta_{1,0}\Delta_{2,0}}{\Delta_{1,0}+\Delta_{2,0}}  K\left(\frac{|\Delta_{1,0}-\Delta_{2,0}|}{\Delta_{1,0}+\Delta_{2,0}}\right)\,.
\label{eq:IJBCS}
\end{equation}
In this appendix, we derive an approximate expression for the critical current (and so the Josephson energy) accounting for the proximity effect in the weak-coupling regime.

We begin by focusing on the integral of $p_1(\e)/(\e+\e')$ over $\e$. Following the approach used for the estimate of the self-consistent order parameter $\dOne$ [cf. Appendix~\ref{app:selfconsistence}], we approximate $p_1(\e)\simeq \mathcal{N}_1(\e)\simeq\beta_1^{1/2}\beta^{1/6}\Im[t_0^{-1}]/2$ (see Appendix~\ref{app:expansionWeak}) for $\tilde\e_g<\e\lesssim\e^*$ and use Eq.~\eqref{eq:p1Approx} for $\e\gtrsim\e^*$, with $\min\{\dTwo-\dOne,\dOne\}\gg\e^*-\dOne\gg \tau_1^{-1}$. The integral of $p_1(\e)/(\e+\e')$ in the region $\tilde\e_g<\e\lesssim \e^*$ at the next-to-leading order in $\beta_j = \tau_j\Delta_j\gg 1$ ($j=1,2$) coincides asymptotically with the integral in the range $\tilde\e_g<\e\lesssim \e^*$ using the simplified BCS pair correlation expression $p_1(\e)\simeq \sqrt{\dOne/2(\e-\tilde\e_g)}$. The analog asymptotic equivalence was explicitly shown for the integral of $p_1(\e)$ in Appendix~\ref{app:selfconsistence} [see Eq.~\eqref{eq:integrationImInvt0} and text that follows it], and can be extended to $p_1(\e)/(\e+\e')$ since at leading order in the expansion the denominator is constant $\e+\e'\simeq \dOne+\e'$ up to linear corrections in $\e-\dOne\leq \e^*-\dOne\ll \dOne+\e'$. As a result, the explicit evaluation of the integral in the two regions yields 
\begin{align}
&\int_{\e_g}^{\e^*}\frac{p_1(\e)}{\e+\e'}d\e\simeq
\sqrt{\frac{2\dOne}{\tilde\e_g+\e'}}\arctan{\sqrt{\frac{\e^*-\tilde\e_g}{\tilde\e_g+\e'}}}\,,
\label{eq:intp1JosEnergyBelowEstar}
\\
&\int_{\e^*}^{+\infty}\frac{p_1(\e)}{\e+\e'}d\e\simeq 
\frac{\dOne}{\sqrt{\e'^2-\dOne^2}}
\Bigg\{
{\rm arccosh}\frac{\e'}{\dOne}
\nonumber\\
&-\log\Bigg[
\left(1+\frac{2\sqrt{\e'^2-\dOne^2}}{\e^*-\sqrt{\e^{*2}-\dOne^2}+\e'-\sqrt{\e'^2-\dOne^2}}\right)
\nonumber\\
&
\left(1-\frac{2\sqrt{\e'^2-\dOne^2}}{\dOne+\e'+\sqrt{\e'^2-\dOne^2}}\right)
\Bigg]
\Bigg\}
+ \frac{\delta(1-\delta)}{\beta_1}f_{\beta_1}(\e^*,\e')\,.
\label{eq:intp1JosEnergyAboveEstar}
\end{align}
The function multiplying the contribution inversely proportional to $\beta_1$ in Eq.~\eqref{eq:intp1JosEnergyAboveEstar} reads
\begin{align}
&f_{\beta_1}(\e^*,\e')=\sqrt{\frac{\dTwo^2 - \e^{*2}}{\e^{*2} - \dOne^2}} \frac{\dOne^2\dTwo^2}{\dTwo^2 - \dOne^2} \frac{\e^{*}/\e' - 1}{\e'^2 - \dOne^2} 
\nonumber\\
&- 
 \frac{\e'^2\dTwo^2}{(\e'^2 - \dOne^2)^{3/2}\sqrt{\e'^2 - \dTwo^2}}\!\arctan\!\sqrt{\frac{(\e'^2 - \dOne^2)(\dTwo^2-\e^{*2})}{(\e'^2 - \dTwo^2)(\e^{*2}-\dOne^2)}}
 \nonumber
 \\
 &+ \frac{\dTwo}{\e'}\Big\{K(\kappa)-F(\phi_1, \kappa)+\frac{\dOne^2}{\e'^2 - \dOne^2} \Pi\left(\frac{\e'^2 \kappa^2}{\e'^2 - \dOne^2},\kappa\right) 
\nonumber\\
&-\frac{\dOne^2}{\e'^2 - \dOne^2}\Pi\left(\frac{\e'^2 \kappa^2}{\e'^2 - \dOne^2},\phi_1, \kappa\right)\Big\}
+\frac{\dOne^2}{\e^{'} \dTwo\kappa^2} \Bigg\{\Pi\left(1, \phi_2, 
\kappa\right)
 \nonumber\\
 &+\frac{\dTwo^2}{\e'^2 - 
      \dOne^2} \left[-E(\phi_3, \kappa)+\frac{\dOne^2}{\dTwo^2}F(\phi_3, \kappa)\right] 
      \Bigg\}
    \nonumber\\&+\frac{\e' \dTwo\dOne^2}{(\e'^2-\dTwo^2)(\e'^2-\dOne^2)} 
 \Pi\left(-\frac{\dTwo^2 \kappa^2}{\e'^2 - \dTwo^2}, \phi_3, \kappa\right)\,,
\label{eq:intp1JosEnergy}
\end{align}
where $\kappa=\sqrt{1-\delta^2}$, $\phi_{1}=\arcsin[\sqrt{\e^{*2}-\dOne^2}/(\e^*\kappa)]$, $\phi_2=\arcsin[\sqrt{\dTwo^2 - \e^{*2}}/(\dTwo\kappa)]$, $\phi_3=\arctan\sqrt{(\dTwo^2 - \e^{*2})/(\e^{*2} - \dOne^2)}$, 
and $\Pi(n,\phi,\kappa)$ is the incomplete elliptic integral of the third kind [we denote with  $\Pi(n,\kappa)= \Pi(n,\pi/2,\kappa)$ the complete one]. Summing Eqs.~\eqref{eq:intp1JosEnergyBelowEstar} and \eqref{eq:intp1JosEnergyAboveEstar}, for $\min\{\dTwo-\dOne,\dOne\}\gg\e^*-\dOne\gg \tau_1^{-1}$, the dependence on $\e^*$ disappears at the relevant order in $1/\beta_1$. In fact, in this limit we replace $\phi_1\to 0$, while $\phi_3$ and the arctan factor in the second line of Eq.~\eqref{eq:intp1JosEnergy} can be replaced with $\pi/2$. Using the expansion (valid for $x\to 0^+$)  
\begin{equation}
\Pi\left(1,\frac{\pi}{2}-x,\kappa\right)\simeq \frac{1}{x\sqrt{1-\kappa^2}}-\frac{E(\kappa)}{1-\kappa^2}+K(\kappa)+O(x)\,,
\end{equation}
in the fourth line of Eq.~\eqref{eq:intp1JosEnergy} [where $\phi_2\simeq \pi/2-\sqrt{2\dOne(\e^*-\dOne)/(\dTwo^2-\dOne^2)}$], and summing to it the contribution in the first line of Eq.~\eqref{eq:intp1JosEnergy}, we obtain in the expansion a term inversely proportional to $\sqrt{\e^*-\dOne}$, which cancels the contribution obtained expanding the right-hand side of Eq.~\eqref{eq:intp1JosEnergyBelowEstar} while the logarithmic term in Eq.~\eqref{eq:intp1JosEnergyAboveEstar} vanishes. Finally, using the connection formula [valid for $(1-n)(1-m)=1-\kappa^2$]~\cite{NIST:DLMF}
\begin{equation}
    (\kappa^2-n)\Pi(n,\kappa)+(\kappa^2-m)\Pi(m,\kappa)=\kappa^2K(\kappa)\,,
\end{equation}
for $n=\e'^2\kappa^2/(\e'^2-\dOne^2)$ and $m=-\dTwo^2\kappa^2/(\e'^2-\dOne^2)$, and rearranging the terms, we get the expression:
\begin{align}
&\int_{\e_g}^{+\infty}\frac{p_1(\e)}{\e+\e'}d\e=\frac{\dOne}{\sqrt{\e'^2-\dOne^2}}
{\rm arccosh}\frac{\e'}{\dOne}
\label{eq:intp1ApproxFinal}\\
&+\frac{1}{\beta_1}\frac{(\dTwo-\dOne)\delta\e'}{\e'^2-\dOne^2}\Bigg\{\frac{-\pi \e'\dTwo}{2\sqrt{(\e'^2-\dTwo^2)(\e'^2-\dOne^2)}}\nonumber\\
&+\frac{1}{\kappa^2}[K(\kappa)-E(\kappa)]+\frac{\dOne^2}{\e'^2-\dOne^2}\Pi\left(\frac{\e'^2\kappa^2}{\e'^2-\dOne^2},\kappa\right)\Bigg\}.
\nonumber
\end{align}
The first term in the right-hand side coincides with what one would obtain by using the BCS expression $p_1(\e)=\dOne\theta(\e^2-\dOne^2)/\sqrt{\e^2-\dOne^2}$. Since the terms in the second and in the third line of Eq.~\eqref{eq:intp1ApproxFinal} are already multiplied by the small factor $\beta_1^{-1}$, the correction associated with this contribution is obtained by using the BCS expression $p_2=\dTwo\theta(\e^2-\dTwo^2)/\sqrt{\e^2-\dTwo^2}$ in Eq.~\eqref{eq:criticalCurrent} and integrating over $\e'$. Using Eq.~\eqref{eq:intp1ApproxFinal}, we immediately find
\begin{align}
    & \int_{\dTwo}^{\infty}\! d\e' \frac{\dTwo}{\sqrt{\e'^2-\dTwo^2}} \int_0^\infty \!d\e\, \Re \frac{p_1(\e)-\frac{\dOne}{\sqrt{\e^2-\dOne^2}}}{\e+\e'} \nonumber \\
    & \simeq (\dTwo-\dOne)\frac{i_1(\delta)}{\beta_1}\,.
\end{align}
where
\begin{align}
i_1(\delta)&=\, \int_{1}^{+\infty} dx\frac{\delta}{\sqrt{x^2-1}}\frac{x}{x^2-\delta^2}\Bigg\{\frac{-\pi x}{2\sqrt{(x^2-1)(x^2-\delta^2)}}\nonumber\\
&+\frac{1}{1-\delta^2}K(\sqrt{1-\delta^2})-\frac{1}{1-\delta^2}E(\sqrt{1-\delta^2})\nonumber\\
&+\frac{\delta^2}{x^2-\delta^2}\Pi\left(\frac{x^2(1-\delta^2)}{x^2-\delta^2},\sqrt{1-\delta^2}\right)
\Bigg\}
\nonumber\\
&=\frac{\pi\, \delta}{2(1-\delta^2)}\left[ \frac{{\rm tanh^{-1}}(\sqrt{1-\delta^2})}{\sqrt{1-\delta^2}}-1\right]
.
\label{eq:correctionEjBeta1}
\end{align}
The expression in the last line of Eq.~\eqref{eq:correctionEjBeta1} is obtained after the variable change $n=x^2(1-\delta^2)/(x^2-\delta^2)$, using the connection formula (with $0<m<1$, and $n\neq 1,m$)~\cite{NIST:DLMF}
\begin{equation}
\Pi(n,\sqrt{m})=K(\sqrt{m})-\Pi(m/n,\sqrt{m})+\Re\left[\frac{\pi\sqrt{n}}{2\sqrt{n-m}\sqrt{1-n}}\right]\,,
\label{eq:connectionFormulaII}
\end{equation}
and the integral
\begin{align}
\int_{m}^{1}dn\sqrt{\frac{n-m}{1-n}}& \Pi(m/n,\sqrt{m})=\pi\bigg\{\frac{3-m}{2}K(\sqrt{m})
\nonumber\\
&-E(\sqrt{m})+\sqrt{m}-{\rm arctanh}\sqrt{m}\bigg\}\,,
\end{align}
which can be proven using the integral representation of $\Pi(m/n,\sqrt{m})$ and swapping the integration order.

As already mentioned, the first term on the right-hand side of Eq.~\eqref{eq:intp1ApproxFinal} corresponds to the result of the integration using the BCS pair correlation for $p_1$. To compute the correction to the Josephson energy which depends on $\beta_2$, it is convenient to first perform the $\e'$ integration of $p_2(\e')/(\e+\e')$. To obtain an approximate expression of the integral, we split the integration range in the three regions (i) $\e_g<\e^\prime\lesssim\e^*$, (ii) $\e^*\lesssim\e^\prime\lesssim\e''$, and (iii) $\e^\prime\gtrsim\e''$ introducing the energies $\e^*, \, \e''$, satisfying the inequalities $ \tau^{-1}_1\ll\e^*-\dOne\ll\min\{\dOne,\dTwo-\dOne\}$ and $\gamma_{2,0}\ll\dTwo-\e''\ll\dTwo-\dOne$ (cf. Appendix~\ref{app:selfconsistence}). As for the case of self-consistent computation of $\dTwo$, 
the contribution in the region (i) is subleading and can be neglected. In region (ii), we use again the leading-order expansion $p_2(\e^\prime)\simeq  p_2^{\rm (ii)}(\e^\prime)=\e^\prime\,\Re[\gamma_2(\e^\prime)]\Delta_2/(\Delta_2^2-\e^{\prime 2})^{3/2}$, with $\gamma_2$ in Eq.~\eqref{eq:explicitGamma_2nd} [assuming $\beta_2 \gg (\beta_1)^{1/2}$]. Direct computation of the integral yields:
\begin{align}
&\int_{\e^*}^{\e^{''}}\!\frac{p_2^{\rm(ii)}(\e')}{\e+\e'}d\e'= \frac{\epsilon(\dTwo-\dOne)}{\beta_2(\epsilon^{2}-\Delta_2^{2})}
\Bigg\{
\Re\Bigg[\frac{\epsilon\dTwo}
{\sqrt{\epsilon^{2}-\Delta_2^{2}} \, \sqrt{\epsilon^{2}-\Delta_1^{2}}}
\nonumber\\
&\times\arctan \Bigg(\frac{-1}{\sqrt{\epsilon^{2}-\Delta_2^{2}}} 
\sqrt{\frac{(\epsilon^{2}-\Delta_1^{2})(\Delta_2^{2}-\epsilon'^{2})}
{\epsilon'^{2}-\Delta_1^{2}}}
\Bigg)\Bigg]
\nonumber\\
&
+\;\frac{\epsilon'/\dTwo - \dTwo/\epsilon}
{\kappa^2}
\sqrt{\frac{\epsilon'^{2}-\Delta_1^{2}}{\Delta_2^{2}-\epsilon'^{2}}}
+\frac{\sqrt{\epsilon'^{2}-\Delta_1^{2}} \, \sqrt{\Delta_2^{2}-\epsilon'^{2}}}
{\epsilon' \dTwo\kappa^{2}}
\nonumber\\
&\quad
-\;\frac{\Delta_1^{2}}{\epsilon^{2}-\Delta_1^{2}} 
\,\Pi \Bigg(
\frac{\epsilon^{2}\kappa^2}{\epsilon^{2}-\Delta_1^{2}}, 
\; \phi_{\e\prime}, 
\;\kappa
\Bigg)
\nonumber\\
&\quad
+\frac{\delta^{2}}{\kappa^{2}} F(\phi_{\e\prime}, \kappa)-\frac{1}{\kappa^{2}} E(\phi_{\e\prime}, \kappa)
\Bigg\}\Bigg|_{\e'=\e^*}^{\e'=\e^{''}}\,,
\label{eq:intp2JosEnergyBelowEstar}
\end{align}
with $\phi_{\e\prime} = \arcsin [
\sqrt{\epsilon'^{2}-\Delta_1^{2}}/(\kappa\epsilon')]$. In region (iii), we approximate $\gamma_{2}\simeq \gamma_{2,0}$ in the parametrization of Eq.~\eqref{eq:pair2Dynes} for $p_2$, and find the following contribution 
\begin{align}
&\int_{\e^{''}}^{+\infty}\frac{p_2^{\rm (iii)}(\e')}{\e+\e'}d\e'= \Re\Bigg\{
\frac{2\dTwo}{\sqrt{(\e-i\gamma_{2,0})^2-\dTwo^2}}\times
\nonumber\\
&{\rm arctanh}\Bigg[\frac{\sqrt{(\e-i\gamma_{2,0})^2-\dTwo^2}}{
\frac{\dTwo^2}{\e^{''}+i\gamma_{2,0}-\sqrt{(\e^{''}+i\gamma_{2,0})^2-\dTwo^2}}+
\e-i\gamma_{2,0}}\Bigg]\Bigg\}
\label{eq:intp2JosEnergyAboveEstar} \\
&\simeq \frac{\dTwo}{\sqrt{\e^2-\dTwo^2}}
{\rm arccosh}\frac{\e}{\dTwo}-\frac{\gamma_{2,0}\sqrt{\dTwo/2}}{(\e+\dTwo)\sqrt{\dTwo-\e^{''}}}
,
\label{eq:intp2JosEnergyAboveEstarApprox} 
\end{align}
where the approximate result in Eq.~\eqref{eq:intp2JosEnergyAboveEstarApprox} is obtained by keeping the first subleading term in the expansion of the right-hand side of Eq.~\eqref{eq:intp2JosEnergyAboveEstar} for $\gamma_{2,0}\ll\dTwo-\e''\ll\dTwo-\dOne$. Summing the contributions from regions (ii) and (iii), and considering the limit for $\gamma_{2,0}\to0$, $\e^*\to\dOne$, $\e''\to\dTwo$ we find the integral up to the linear order in $\beta_2^{-1}$. In particular, $\phi_{\epsilon^*}\to 0$ as $\e^*\to\dOne$ and the terms in the last three lines in Eq.~\eqref{eq:intp2JosEnergyBelowEstar} tend to zero, in agreement with neglecting the contribution from region (i), while the arctan function in the second line can be replaced with $-\pi\theta(\e^2-\dTwo^2)/2$ . For $\e''\to \dTwo$, the elliptic integrals in Eq.~\eqref{eq:intp2JosEnergyBelowEstar} become complete (since $\phi_{\e^{''}}\to\pi/2$), and the argument of the arctangent tends to zero. The residual dependence on $\e''$  due to the third line of Eq.~\eqref{eq:intp2JosEnergyBelowEstar} disappears, since the left term in this line is compensated by the last term in Eq.~\eqref{eq:intp2JosEnergyAboveEstarApprox} while the right term vanishes.
In summary, we have 
\begin{align}
&\int_{\e_g}^{+\infty}\frac{p_2(\e')}{\e+\e'}d\e'\simeq\frac{\dTwo}{\sqrt{\e^2-\dTwo^2}}
{\rm arccosh}\frac{\e}{\dTwo}
\label{eq:intp2ApproxFinal}\\
&+\frac{1}{\beta_2}\frac{\e(\dTwo-\dOne)}{\e^2-\dTwo^2}\Bigg\{\frac{\pi \e\dTwo\theta(\e^2-\dTwo^2)}{2\sqrt{(\e^2-\dTwo^2)(\e^2-\dOne^2)}}\nonumber\\
&+\frac{1}{\kappa^2}[\delta^2 K(\kappa)-E(\kappa)]-\frac{\dOne^2}{\e^2-\dOne^2}\Pi\left(\frac{\e^2\kappa^2}{\e^2-\dOne^2},\kappa\right)\Bigg\}.
\nonumber
\end{align}
Using Eq.~\eqref{eq:intp2ApproxFinal}, we find
\begin{align}
    & \int_{\dOne}^{\infty}\!d\e  \frac{\dOne}{\sqrt{\e^2-\dOne^2}} \int_0^\infty\! d\e' \, \Re\frac{p_2(\e')-\frac{\dTwo}{\sqrt{\e'^2-\dTwo^2}}}{\e+\e'} \nonumber \\ & \simeq (\dTwo-\dOne)\frac{ i_2(\delta)}{\beta_2}\,,
\end{align}
where
\begin{align}
i_2(\delta)&=\int_{\delta}^{+\infty} dx\frac{\delta}{\sqrt{x^2-\delta^2}}\frac{x}{x^2-1}\Bigg\{\frac{\pi x \theta(x-1)}{2\sqrt{(x^2-1)(x^2-\delta^2)}}\nonumber\\
&+\frac{\delta^2}{1-\delta^2}K(\sqrt{1-\delta^2})-\frac{1}{1-\delta^2}E(\sqrt{1-\delta^2})\nonumber\\
&-\frac{\delta^2}{x^2-\delta^2}\Pi\left(\frac{x^2(1-\delta^2)}{x^2-\delta^2},\sqrt{1-\delta^2}\right)
\Bigg\}\nonumber\\
&=\frac{\pi\delta}{2(1-\delta^2)}\left[\frac{\delta\arcsin{\sqrt{1-\delta^2}}}{\sqrt{1-\delta^2}}-1\right]
.
~\label{eq:correctionEjBeta2}
\end{align}
The expression in the last line of Eq.~\eqref{eq:correctionEjBeta2} is obtained by splitting the integration range in the regions (a) $\delta\leq x\leq1$ and (b) $x\geq 1$, with contributions $i_2^{(a)}(\delta)$, and $i_2^{(b)}(\delta)$. For both the terms, we change the integration variable $n=x^2(1-\delta^2)/(x^2-\delta^2)$, and use again the transformation in Eq.~\eqref{eq:connectionFormulaII} for $i_2^{(b)}(\delta)$. Using the integral representation of the elliptic function of the third kind, after some algebra we find 
\begin{align}
i_2^{(a)}(\delta)&= \frac{\pi\delta[\delta\arcsin\sqrt{1-\delta^2}-\sqrt{1-\delta^2}]}{2(1-\delta^2)^{3/2}}-i_2^{(b)}(\delta)\,,\\
i_2^{(b)}(\delta)&=\!\int_{0}^{2\pi}\!\!\!d\theta \frac{\delta^2\sin(\theta)\sin(2\theta)\arctan[\delta\sec(\theta)/\sqrt{1-\delta^2}]}{-2[1-\sqrt{1-\delta^2}\sin(\theta)^2]^{3/2}},
\end{align}
from which we immediately find the desired expression. Combining the results of Eqs.~\eqref{eq:correctionEjBeta1} and~\eqref{eq:correctionEjBeta2}, we can write the renormalized Josephson energy as
\begin{align}
   E_J&=\frac{g_T}{e^2}\frac{\Delta_{1}\Delta_{2}}{\Delta_{1}+\Delta_{2}}  K\left(\frac{|\Delta_{1}-\Delta_{2}|}{\Delta_{1}+\Delta_{2}}\right)
   \nonumber\\&+
   \frac{g_T(\dTwo-\dOne)}{\pi e^2}\left[\frac{i_1(\delta)}{\beta_1}-\frac{i_1(1/\delta)}{\beta_2}\right].
\end{align}

The separation of the leading correction into two terms suggest that it should be possible to treat separately the two asymptotic limits $\beta_1 \gg 1$ with $\beta_2 \to \infty$ and $\beta_2 \gg 1$ with $\beta_1 \to \infty$. These limits are more easily studied using the single-integral formula for the critical current $I_J$~\cite{LO1966}
\begin{equation}
    I_J = \frac{g_T}{e} \int_0^{+\infty} d\epsilon \,\Im \left[\sqrt{1+X_1^2(\epsilon)}\sqrt{1+X_2^2(\epsilon)}\right] .
\end{equation}
This expression can be derived in a straightforward way from Eq.~\eqref{eq:criticalCurrent} by multiplying both the numerator and the denominator in the integral by $\e-\e'$, taking the Cauchy principal value, and using the Kramers-Kronig relation (we remind that $p_j=\Im[\sqrt{1+X_j^2}]$). When $\beta_2 \to \infty$, the $X_2$ factor takes the BCS form $\sqrt{1+X_2^2}= \dTwo/\sqrt{\dTwo^2-\epsilon^2}$ and the $X_1$ one is given by $\sqrt{1+X_1^2}=(1+u^2)/2u$ with $u$ in Eq.~\eqref{eq:ub1exact}. Analogously, when $\beta_1 \to \infty$, the $X_1$ factor takes the BCS form $\sqrt{1+X_1^2}=\dOne/\sqrt{\dOne^2-\epsilon^2}$ and the $X_2$ one can be written in the Dynes form $\sqrt{1+X_2^2}=\dTwo/\sqrt{\dTwo^2-[\epsilon+i\gamma_2(\epsilon)]^2}$ with $\gamma_2$ as in Eq.~\eqref{eq:explicitGamma}. While the resulting integrals are not directly amenable to perturbative treatment, they can be rewritten in a more useful form by completing the integration contour in the upper right quadrant of the complex plane; then, in addition to the original integral, only the portion along the imaginary axis contributes. Effectively, one can replace $\epsilon \to i\epsilon$ and start the integration from 0. In this form, the integrals can be expanded perturbatively in the small parameters $1/\beta_j$; after integration, the prefactors obtained for the corrections agree with those calculated above.

\section{Analytical approximations for spectral density and frequency shift}
\label{app:detailsApprox}
In this Appendix, we provide details on the derivation of the analytical approximations for the spectral density entering the qubit transition rate [cf. Eq.~\eqref{eq:Gamma10}] and for the regularization of the jump in the frequency shift, Eq.~\eqref{eq:jprox}.
\subsection{Spectral density}
For the calculation of the spectral density $S_1(\omega)$, we approximate $f(\e)\simeq e^{-\e/T}$, $1-f(\e+\omega)\simeq 1$ since $T\ll \epsilon_g$, and the coherence factor with $F_{12}(\e,\e+\w)\simeq \{1+\dOne\dTwo/[\e_g(\e_g+\w)]\}/2\approx (\dOne+\dTwo+\w)/[2(\dOne+\w)]$, under the assumption $\w \gg \sqrt{(\dTwo-\dOne)/(\dTwo+\dOne)}\big/\tau_1$ [cf. Eq.~\eqref{eq:conditionGammaExplicit}]. Thus, the spectral function approximately reads:
\begin{equation}
S_1(\w)=\frac{2g_T}{e^2}\frac{\dOne+\dTwo+\w}{\dOne+\w}\int_{\e_g}^{\infty}\!\!d\e \, \mathcal{N}_1(\e)
\mathcal{N}_2(\e+\w)e^{-\e/T}.
\label{eq:SqpPlus}
\end{equation}
Since the proximity coupling mainly affects the rates close to the resonant condition $\wLR=\omega_{01}$, we restrict our analysis to frequencies $\omega\simeq\wLR$. We split the integration range into the two regions (i) $\e_g<\e \lesssim \e_g+\xi$ and (ii) $\e\gtrsim \e_g+\xi$, where $\xi=c\tilde\gamma_1$ with $c$ a numerical factor of order unity (its exact value is irrelevant within our approximations). In region (i), we approximate $f(\e)\simeq e^{-\e_g/T}$ for $T \gg \tilde{\gamma}_1$; the width of the integration region is of order $\tilde{\gamma}_1\approx \dOne/(\tau_1\dOne\tau_2\dTwo)^{2/3}$, and $\mathcal{N}_1$ and $\mathcal{N}_2$ are bounded by factors of order $(\tau_1\dOne\tau_2\dTwo)^{1/3}$ and $(\tau_2\Delta_2)^{1/2}$, respectively (see Sec.~\ref{sec:approxSolutions}). Therefore, we estimate the contribution from this region to be of order $\dOne e^{-\e_g/T}(\tau_2\dTwo)^{1/6}/(\tau_1\dOne)^{1/3}$. Assuming $\tau_1\dOne \gtrsim (\tau_2\Delta_2)^{1/2}$ (more precisely, $\gamma_{2,0} \gtrsim\tilde\gamma_1$), this contribution can be neglected 
compared to the contribution from region (ii), which returns Eq.~\eqref{eq:S_1PlusRelaxationExplicit} as we show next. 
Let us note that the assumption made here is not in contradiction with that made in Sec.~\ref{sec:lowerGapDoS}, namely $\tau_1\dOne \ll (\tau_2\Delta_2)^{2}$; the two assumptions restrict the range of values for $\tau_1\dOne$, but since typically in experiments with thin films $\tau_1\dOne\sim \tau_2\dTwo$, we do not consider here other limiting cases. In region (ii) we approximate $\mathcal{N}_1(\e)\simeq\sqrt{\Delta_1/2(\e-\tilde\e_g)}$  and $\mathcal{N}_2(\e+\w)\simeq \sqrt{\Delta_2/2}\,\Re[1/\sqrt{\e+\w+i\gamma_{2,0}-\dTwo}]$ obtained using Eq.~\eqref{eq:explicitGamma_2nd} (with $\e=\dTwo$) in Eq.~\eqref{eq:DynesAnsatz} ($j=2$); this approximation for $\mathcal{N}_2$ is accurate for our goals, since the main contribution to the integral of Eq.~\eqref{eq:SqpPlus} stems from energies $\e\gtrsim \e_g$, such that $\e+\w\simeq \dTwo$. With these approximations in Eq.~\eqref{eq:SqpPlus} we find Eq.~\eqref{eq:S_1PlusRelaxationExplicit} for the spectral function.
\subsection{Frequency shift}
Close to the resonant condition, $\w\simeq\omega_{\delta\Delta}$, and for $T\ll \omega_{\delta\Delta}$, Eq.~\eqref{eq:ImYqpBCS} approximately reads:
\begin{align}\label{eq:ImYqpBCSRes}
 \Im[{Y_{\rm qp,BCS}^{\rm dyn}}] \simeq
\frac{\pi g_T\bar\Delta}{2\w_{\delta\Delta}}e^{\frac{-\dOne}{T}}\Bigg[1&-h(\omega-\omega_{\delta\Delta}) 
\nonumber
\\ &
-(4-\sqrt{2})\sqrt{\frac{T}{\pi \omega_{\delta\Delta}}}\Bigg],
\end{align}
where we used the leading-order asymptotic expansions $e^{\pm z}I_0(z)\simeq 1$ for $z\to 0$ and $I_0(z) \simeq e^z/\sqrt{2\pi z}$ for $z\to \infty$.
Due to the term $h(\omega-\omega_{\delta\Delta})$, $\Im[{Y_{\rm qp}^{\rm dyn}}]$ displays a jump discontinuity at $\w=\dTwo-\dOne$, which is related to the divergence of the spectral density in Eq.~\eqref{eq:S_1PlusBCS}. When proximity effects are included the logarithmic divergence is regularized, and so the jump is replaced by a continuous function. This regularization can be captured by replacing $h(\omega)$ with a smooth function $h_{\rm prox}(\omega,\gamma_{2,0})$, which must reduce to the BCS result in the limit $\tau_i\to \infty$, $\lim_{\gamma_{2,0}\to 0^+}h_{\rm prox}(\omega,\gamma_{2,0})={\rm sign}(\omega)$. To determine $h_\mathrm{prox}$ we proceed as follows:
close to the resonance, $|\omega-\omega_{\delta\Delta}|\ll \gamma_{2,0}\ll T\ll\omega,\omega_{\delta\Delta}$,
we approximate $S(\omega)-S(-\omega)\simeq S_1(\omega)$ in Eq.~\eqref{eq:ReYqp} and use for the latter the approximate expression of Eq.~\eqref{eq:regularizedDiv} [where we also approximate $\sqrt{\dOne\dTwo}\simeq\bar\Delta]$; then, using the standard result for the Hilbert transform (see, for instance, Ref.~\cite{bookHilbertTransform}):
\begin{align}
\label{eq:hilbertTransformLog}
& \mathcal{H}\left[\frac{\ln[(\omega-\omega_{\delta\Delta})^2+\gamma_{2,0}^2]}{\omega}\right]=  \\
&-\frac{2}{\omega}\{\arctan\left[(\omega-\omega_{\delta\Delta})/\gamma_{2,0}\right]
+\arctan\left[\omega_{\delta\Delta}/\gamma_{2,0}\right]\}\,, \nonumber
\end{align}
from Eq.~\eqref{eq:ImYqpintegral} we obtain (approximating $\tilde\e_g\simeq\dOne$)
\begin{equation}
\Im[{Y_{\rm qp}^{\rm dyn}}]\simeq
-\frac{\pi g_T\bar\Delta}{2\w_{\delta\Delta}}e^{\frac{-\dOne}{T}}\left[1+\frac{2}{\pi}\arctan\left(\frac{\omega-\omega_{\delta\Delta}}{\gamma_{2,0}}\right)\right]
\,,
\label{eq:ImYqpProx}
\end{equation}
where we approximated the second line of Eq.~\eqref{eq:hilbertTransformLog} for $\gamma_{2,0}\ll \omega_{\delta\Delta}$.
Comparing the rapidly varying terms in
Eqs.~\eqref{eq:ImYqpBCSRes} and \eqref{eq:ImYqpProx}, we immediately 
find the function quoted in the main text, Eq.~\eqref{eq:jprox}. In closing this Appendix, we note that Eq.~\eqref{eq:ImYqpProx} cannot capture the $\omega$-independent terms in Eq.~\eqref{eq:ImYqpBCSRes} for $\gamma_{2,0}\to 0$: the procedure used disregards such terms, as the Hilbert transform of a constant vanishes.

\section{Strong coupling}
\label{app:strong}
In this Appendix, we briefly discuss the gap of the superconducting bilayer for small interface resistance between the two layers, i.e., $\tau_1\dOne,\tau_2\dTwo\ll 1$. Under the condition 
\begin{equation}
\left|\tau_j^2(\Delta_j^2-\e^2)+2\tau_j\left[\Delta_j\sqrt{X_{\tilde{j}}^2+1}-\e X_{\tilde{j}}
\right]
\right|\ll 1
\end{equation}
we can approximate to unity the term in the second line of Eq.~\eqref{eq:selfXj}, and solve for $X_j$; the solution is BCS-like  $X_j=\epsilon/\sqrt{\tilde\epsilon_{g,\rm sc}^2-\epsilon^2}$, with a gap 
\begin{equation}
\tilde\epsilon_{g,\rm sc}=\frac{\tau_1\dOne+\tau_2\dTwo}{\tau_1+\tau_2}. 
\label{eq:egTildeSC}
\end{equation}
This result was derived in the seminal work of McMillan~\cite{McMillanPR175} within a tunneling model (see Ref.~\cite{Note2}), and, for $\dOne=0$, this result correctly reduces to the one reported in Refs.~\cite{FominovPRB63,HosseinkhaniPRB97} for NS bilayers in the strong coupling regime.
To perform the calculation at the next to leading order, we use again the $u$ parametrization (see Appendix~\ref{app:expansionWeak}). We start by conveniently rewriting Eq.~\eqref{eq:UsadelScaledU} as
\begin{align}
&u(\beta_1^2-\delta^2\beta_2^2)
[\bar\epsilon_{g,{\rm sc}}-\tilde\epsilon -u^2 (\bar\epsilon_{g,{\rm sc}}+\tilde\epsilon)]
[\tilde\epsilon_0-\tilde\epsilon -u^2 (\tilde\epsilon_0+\tilde\epsilon)]
\nonumber\\
&= -\beta_1^2\beta_2[1-\delta\tilde\epsilon+ u\beta_2(1-\delta^2\tilde\epsilon^2)+(1+\delta\tilde\epsilon)u^2]
\nonumber\\ &\qquad \times [1-\tilde\epsilon-u^2(1+\tilde\epsilon)]^2,
\label{eq:UstrongCoupling}
\end{align}
with $\bar\epsilon_{g,\mathrm{sc}}=\tilde\epsilon_{g,\mathrm{sc}}/\dOne$, and $\tilde\epsilon_0=(\beta_1-\beta_2)/(\beta_1-\delta\beta_2)$~\footnote{Note that $\tilde\epsilon_0$ diverges for $\beta_1-\delta\beta_2=0$, i.e., $\tau_1=\tau_2$ or $\dOne=0$. In this case, the prefactor collected on the left-hand side regularizes the expression. This detail doesn't change our discussion, since the final result behaves well both for $\tau_1=\tau_2$ and $\dOne=0$.}. 
Since at the leading order $X_j$ is approximately BCS-like with gap $\tilde\epsilon_{g,\mathrm{sc}}$, we have $u=0$ for $\tilde\epsilon=\bar\epsilon_{g,{\rm sc}}$ at this order of the perturbative expansion,   as for the weak-coupling case (see Appendix~\ref{app:expansionWeak}). In order to compute the next-to-leading order solution, we apply the dominant balance analysis to Eq.~\eqref{eq:UstrongCoupling} for $\tilde\epsilon=\bar\epsilon_{g,{\rm sc}}$, which returns a scaling $u\propto \beta_2^{1/3}$ for $\beta_1,\beta_2\ll 1$. Specifically, we have $u\ll 1$ for $\tilde\epsilon=\bar\epsilon_{g,{\rm sc}}$ and the dominant term in the left-hand-side of Eq.~\eqref{eq:UstrongCoupling}, obtained setting $u=0$ in the rightmost square parenthesis, balances the term $-\beta_1^2\beta_2(1-\delta\bar{\e}_{g,{\rm sc}})(1-\bar{\e}_{g,{\rm sc}})^2$ in the right-hand side of the same equation.  

To compute the next-to-leading order correction to the gap, we keep terms of order $\beta$ in Eq.~\eqref{eq:UstrongCoupling}, assuming $|\tilde\epsilon-\bar\epsilon_{g,{\rm sc}}|\sim \beta_2^{2/3}\ll 1$. We obtain a depressed cubic equation
\begin{equation}
u^3+\left(\frac{\epsilon}{\tilde\epsilon_{g,{\rm sc}}}-1\right)\frac{u}{2} 
+ \frac{1}{4(\beta_1+\beta_2)}\left[\frac{\tau_1\tau_2(\dTwo-\dOne)}{\tau_1+\tau_2}\right]^2=0 \, ,
\end{equation}
and the correction to the gap is found setting the cubic discriminant to zero, reading
\begin{equation}
\epsilon_g\approx\tilde\epsilon_{g,{\rm sc}}
\left[
1-\frac{3}{2}\frac{\tau_1^{4/3}\tau_2^{4/3}(\dTwo-\dOne)^{4/3}}{(\tau_1+\tau_2)^{4/3}}\frac{1}{(\beta_1+\beta_2)^{2/3}}
\right].
\label{eq:minigapStrongNext}
\end{equation}
This expression for $\dOne=0$ correctly reproduces the next-to-leading order correction to the gap for the NS bilayer derived in Ref.~\cite{HosseinkhaniPRB97}.
\subsection{Approximations for the self-consistent order parameters}
Here we discuss some approximations for the order parameters in the strong coupling regime $\tau_j\Delta_j\ll 1$. To simplify the analytical calculations, we assume $\omega_{D,1}=\omega_{D,2}=\omega_D$ in this section. The approach presented here extends the one given in Ref.~\cite{FominovPRB63} for NS bilayers.
In the strong coupling regime, the asymptotic behavior for $\e\gg \dOne,\dTwo$ of the pair correlation function is modified by the proximity coupling. This feature can be observed by linearizing the Usadel equations Eq.~\eqref{eq:UsadelAngular} in this limit, which yields a solution for the pairing angles:
\begin{equation}
\theta_j(\e)\approx i\frac{\Delta_j}{\e}\frac{\tilde{\e}_{g,\mathrm{sc}}/\Delta_j-i\bar\tau\e}{1-i\bar\tau\e}\, ,
\label{eq:thetajAsymptotic}
\end{equation}
where we introduced the harmonic mean of $\tau_1,\tau_2$, i.e.,
\begin{equation}
\bar{\tau}=\frac{\tau_1\tau_2}{\tau_1+\tau_2}\,.
\end{equation}
The asymptotic solutions of Eq.~\eqref{eq:thetajAsymptotic} hold for arbitrary values of the proximity coupling, as discussed in Ref.~\cite{FominovPRB63}. Indeed, in the weak coupling regime, we have $\bar\tau\e\gg 1$ and so the asymptotic solution reduced to the uncoupled case, as expected. To evaluate the integral in the self-consistent equation, we introduce a characteristic energy $\e^{**}$, requiring that $1/\bar\tau \gg \e^{**}\gg \dTwo$. For $\e\leq \e^{**}$, the pairing angle can be expressed as $\theta_j(\e)=\arctan[i\tilde\e_{g,\mathrm{sc}}/\e]$, corresponding to the BCS-like solution with gap $\tilde\e_{g,\mathrm{sc}}$ (see discussion in this Appendix), while at higher energies can be approximated using Eq.~\eqref{eq:thetajAsymptotic}. The asymptotic value of the integral is independent on $\e^{**}$, and reads
\begin{equation}
\int_0^{\omega_D}p_jd\e= \tilde\e_{g,\mathrm{sc}}\ln\left(\frac{2\omega_D}{\tilde\e_{g,\mathrm{sc}}}\right)
+(\Delta_j-\tilde\e_{g,\mathrm{sc}})\ln\sqrt{1+\bar\tau^2\omega_D^2}\,,
\end{equation}
and, once combined with the BCS self-consistent equation, return the implicit equation
\begin{equation}
\frac{\tilde\e_{g,\mathrm{sc}}}{\Delta_{j,0}}=\left(\frac{\Delta_{j,0}}{2\omega_D}\sqrt{1+\bar\tau^2\omega_D^2}\right)^{\Delta_j/\tilde\e_{g,\mathrm{sc}}-1}\,.
\label{eq:implicitOrderParameterSC}
\end{equation}

The solution to Eq.~\eqref{eq:implicitOrderParameterSC} can be found in a closed form. First, we rewrite that equation as
\begin{equation}
\frac{\sqrt{1+\bar\tau^2\omega_D^2}}{2\omega_D}\tilde\e_{g,\mathrm{sc}}=\left(\frac{\Delta_{j,0}}{2\omega_D}\sqrt{1+\bar\tau^2\omega_D^2}\right)^{\Delta_j/\tilde\e_{g,\mathrm{sc}}}\,.
\label{eq:implicitOrderParameterSC2}
\end{equation}
and we take the natural logarithm of the two equations with $j=1,\,2$; this enables us to relate $\dOne$ and $\dTwo$
\begin{equation}
\dOne \ln\tilde\Delta_{1,0} = \dTwo \ln\tilde\Delta_{2,0}\,,
\label{eq:D1D2relationSC}
\end{equation}
where we introduce the short-hand notation
\begin{equation}
\tilde\Delta_{j,0}=\Delta_{j,0}\frac{\sqrt{1+\bar\tau^2\omega_D^2}}{2\omega_D}\,.
\end{equation}
Combining Eqs.~\eqref{eq:egTildeSC} and \eqref{eq:D1D2relationSC}, we can rewrite the exponent in the right-hand side of Eq.~\eqref{eq:implicitOrderParameterSC2} as
\begin{equation}
\label{eq:DeltajegTildeRatioSC}
    \frac{\Delta_j}{\tilde\e_{g,\mathrm{sc}}}=\frac{\tau_1+\tau_2}{\tau_j+\tau_{\bar j}\Delta_{\bar{j}}/\Delta_{j}} = \frac{\tau_1+\tau_2}{\tau_j+\tau_{\bar j}\ln\tilde\Delta_{j,0}/\ln\tilde\Delta_{\bar{j},0}}\,.
\end{equation}
Substituting the last formula into Eq.~\eqref{eq:implicitOrderParameterSC2}, we can express $\tilde\e_{g,\mathrm{sc}}$ in terms of $\Delta_{1,0}$, $\Delta_{2,0}$, $\bar\tau$ and $\omega_D$. Finally, using such expression in Eq.~\eqref{eq:DeltajegTildeRatioSC}, we find
\begin{align}
\Delta_j & =
\frac{2\omega_D}{\sqrt{1+\bar\tau^2\omega_D^2}}
\frac{\ln(\tilde\Delta_{\bar j,0})(\tau_1+\tau_2)}{\tau_1\ln(\tilde\Delta_{2,0})+\tau_2\ln(\tilde\Delta_{1,0})}
\nonumber \\ & \times e^{\frac{(\tau_1+\tau_2)\ln(\tilde\Delta_{1,0})\ln(\tilde\Delta_{2,0})}{\tau_1\ln(\tilde\Delta_{2,0})+\tau_2\ln(\tilde\Delta_{1,0})}}\,.
\label{eq:DeltaApproxSCstrong}
\end{align}

\begin{figure}[!tb]
    \centering
\includegraphics[width=0.42\textwidth]{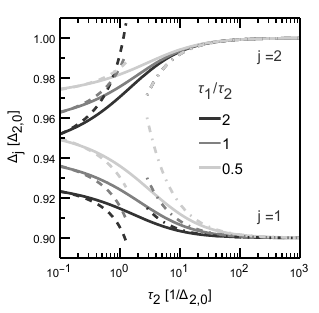}
    \caption{Self-consistent order parameters $\Delta_j$ vs $\tau_2$ for $\Delta_{1,0}=0.9\Delta_{2,0}$ and $\tau_1/\tau_2=2,\, 1,\, 0.5$ (dark to light gray). Solid lines are computed solving numerically the gap equation in the Matsubara representation (see text). The dot-dashed curves are the analytical approximations in Eqs.~\eqref{eq:Delta1ApproxSCweak} and \eqref{eq:Delta2ApproxSCweak} for the weak-coupling regime, while the dashed curves are the approximation in Eq.~\eqref{eq:DeltaApproxSCstrong} for the strong-coupling regime (in the limit $\bar\tau\omega_{D}\to\infty $).
    }
    \label{fig:Figure8}
\end{figure}

For $\dOne=0$ (NS bilayer), the results given in Ref.~\cite{FominovPRB63} can be easily reproduced using Eq.~\eqref{eq:implicitOrderParameterSC} and the first equality in Eq.~\eqref{eq:DeltajegTildeRatioSC}; in particular, the latter gives $\dTwo/\tilde\e_{g,sc}-1=\tau_1/\tau_2$.

Figure~\ref{fig:Figure8} displays the self-consistent order parameters $\Delta_j$ as a function of $\tau_2\Delta_{2,0}$ for $\Delta_{1,0}=0.9\Delta_{2,0}$ and few values of the ratio $\tau_1/\tau_2$ (grayscale). The solid curves in the figure are obtained numerically, following the method described at the beginning of Appendix~\ref{app:selfconsistence}. As expected, the order parameters approach the values they have in the absence of the proximity effect as $\tau_i\to \infty$. By lowering $\tau_2\Delta_{2,0}$, the difference between the two order parameters monotonically decreases. Our approximations for weak coupling, Eqs.~\eqref{eq:Delta1ApproxSCweak} and \eqref{eq:Delta2ApproxSCweak} (dot-dashed), and for strong coupling, Eq.~\eqref{eq:DeltaApproxSCstrong} (dashed), accurately describe the values computed numerically in the appropriate limits. 
\textcolor{white}{\begin{NoHyper}\cite{Womak2021}\end{NoHyper}}
\bibliography{references}
\end{document}